\documentclass[journal,compsoc]{IEEEtran}

\usepackage{lipsum}
\usepackage{graphicx,wrapfig,lipsum}

\usepackage{amsmath}
\usepackage{etoolbox,siunitx}
\robustify\bfseries
\usepackage{booktabs} 
\usepackage{enumitem} 
\usepackage[mathscr]{euscript}
\usepackage{graphicx}
\usepackage{graphics}
\usepackage{setspace}
\usepackage{url}
\usepackage{amsmath,amssymb}
\usepackage{amsfonts}
\usepackage{dsfont}
\usepackage{epsfig}
\usepackage{algorithm}
\usepackage{algorithmic}
\usepackage{rotating}
\usepackage{adjustbox}
\usepackage{array,multirow}
\usepackage{hhline}
\usepackage[justification=centering]{caption}
\usepackage{wrapfig}
\usepackage{float}
\usepackage{siunitx}
\usepackage{xspace} 
\usepackage{tabularx} 
\graphicspath{ {img/} }
\usepackage{tikz}

\newcolumntype{R}[2]{%
    >{\adjustbox{angle=#1,lap=\width-(#2)}\bgroup}%
    l%
    <{\egroup}%
}

\makeatletter
\newcommand\fs@nobottomruled{\def\@fs@cfont{\bfseries}\let\@fs@capt\floatc@ruled
  \def\@fs@pre{\vspace{-0mm}\hrule height.8pt depth0pt\kern0.5pt}%
  \def\@fs@post{\kern0.5pt\hrule\relax\vspace{-4mm}}%
  \def\@fs@mid{\kern0.5pt\hrule\kern0.5pt}%
  \let\@fs@iftopcapt\iftrue}
\makeatother

\floatstyle{nobottomruled}
\restylefloat{algorithm}

\newcommand{\spara}[1]{\vspace{0.1cm}\noindent{\bf #1}}

\newtheorem{definition}{Definition}

 \newtheorem{theorem}{Theorem}
 \newtheorem{corollary}{Corollary}

 \newtheorem{problem}{Problem}

\DeclareMathOperator*{\argmax}{arg\,max}

\newcommand{\NPhard}{$\mathbf{NP}$-hard\xspace}

\renewcommand{\vec}[1]{\mathbf{#1}}

\newcommand{\squishlist}{
 \begin{list}{$\bullet$}
  {  \setlength{\itemsep}{0pt}
     \setlength{\parsep}{3pt}
     \setlength{\topsep}{3pt}
     \setlength{\partopsep}{0pt}
     \setlength{\leftmargin}{2em}
     \setlength{\labelwidth}{1.5em}
     \setlength{\labelsep}{0.5em}
} }
\newcommand{\squishlisttight}{
 \begin{list}{$\bullet$}
  { \setlength{\itemsep}{0pt}
    \setlength{\parsep}{0pt}
    \setlength{\topsep}{0pt}
    \setlength{\partopsep}{0pt}
    \setlength{\leftmargin}{2em}
    \setlength{\labelwidth}{1.5em}
    \setlength{\labelsep}{0.5em}
} }

\newcommand{\squishdesc}{
 \begin{list}{}
  {  \setlength{\itemsep}{0pt}
     \setlength{\parsep}{3pt}
     \setlength{\topsep}{3pt}
     \setlength{\partopsep}{0pt}
     \setlength{\leftmargin}{1em}
     \setlength{\labelwidth}{1.5em}
     \setlength{\labelsep}{0.5em}
} }

\newcommand{\squishend}{
  \end{list}
}

\newcommand{\f}{\ensuremath{f}\xspace}

\newcommand{\bigO}{\mathcal{O}\xspace}

\newcommand{\unicredit}{\textsf{UniCredit}\xspace}
\newcommand{\amount}{\ensuremath{amount}\xspace}
\newcommand{\creditor}{\ensuremath{creditor}\xspace}
\newcommand{\debtor}{\ensuremath{debtor}\xspace}

\newcommand{\insertdate}{\ensuremath{indate}\xspace}
\newcommand{\duedate}{\ensuremath{duedate}\xspace}
\newcommand{\lifetime}{\ensuremath{life}\xspace}

\newcommand{\balr}{\ensuremath{bl_r}\xspace}
\newcommand{\bala}{\ensuremath{bl_a}\xspace}

\newcommand{\ub}{\ensuremath{cap}\xspace}
\newcommand{\lb}{\ensuremath{fl}\xspace}

\newcommand{\staticbalance}{\textsc{Max-profit Balanced Settlement}\xspace}
\newcommand{\rstaticbalance}{\textsc{Relaxed Settlement}\xspace}

\newcommand{\mincostflow}{\textsc{Min-Cost Flow}\xspace}

\newcommand{\subsetsum}{\mbox{\textsc{Subset Sum}}\xspace}
\newcommand{\bbalg}{\mbox{\textsf{Settle-\textsc{bb}}}\xspace}
\newcommand{\optcyclesel}{\textsc{Optimal Cycle Selection}\xspace}
\newcommand{\maxindset}{\textsc{Maximum Independent Set}\xspace}
\newcommand{\johnson}{\textsf{Settle-\textsc{bb-lb}}\xspace}

\newcommand{\wsetcover}{\mbox{\textsc{Weighted Set Cover}}\xspace}
\newcommand{\maxcyclelength}{\ensuremath{L}\xspace}
\newcommand{\flowgraph}{\textsf{R}-flow graph\xspace}
\newcommand{\bbalgub}{\textsf{\bbalg-\textsc{ub}}\xspace}
\newcommand{\bmultigraph}{\textsf{R}-multigraph\xspace}
\newcommand{\suk}{\textsc{Set Union Knapsack}\xspace}
\newcommand{\multisuk}{\textsc{Multidimensional Set Union Knapsack}\xspace}
\newcommand{\cycles}{\ensuremath{\mathscr{C}}\xspace}
\newcommand{\paths}{\ensuremath{\mathscr{P}}\xspace}
\newcommand{\bsalg}{\mbox{\textsf{Settle-\textsc{beam}}}\xspace}
\newcommand{\knapsack}{\textsc{Knapsack}\xspace}
\newcommand{\cyclescoreSUK}{\ensuremath{\omega}\xspace}
\newcommand{\maxcover}{\textsc{Max Cover}\xspace}
\newcommand{\hybridalg}{\textsf{Settle-\textsc{h}}\xspace}
\newcommand{\hybridalgp}{\textsf{Settle-\textsc{h-path}}\xspace}
\newcommand{\pathalg}{\textsf{Settle-\textsc{path}}\xspace}
\newcommand{\pathalggreedy}{\textsf{Settle-\textsc{h-path}-\textsc{g}}\xspace}
\newcommand{\pathalgsuk}{\textsf{Settle-\textsc{h-path}-\textsc{s}}\xspace}
\newcommand{\rfbaseline}{\textsf{RFB}\xspace}
\newcommand{\negbalalg}{\textsf{Redefine-Floors}\xspace}
\newcommand{\negbalalgtwo}{\textsf{Select-and-Order}\xspace}
\newcommand{\firstlog}{\textsf{\ensuremath{Log_{15}}}\xspace}
\newcommand{\secondlog}{\textsf{\ensuremath{Log_{17}}}\xspace}
   {\vspace*{-2pt}
    \begin{enumerate}[wide=3pt, leftmargin=\dimexpr\labelwidth + 2\labelsep\relax]}%
   {\end{enumerate}
    \vspace*{-2pt}}

\newenvironment{smallitemize}%
   {\vspace*{-2pt}
    \begin{itemize}[wide=3pt, leftmargin=\dimexpr\labelwidth + 2\labelsep\relax]}%
   {\end{itemize}
    \vspace*{-2pt}}

\newcommand{\G}{\ensuremath{\mathcal{G}}}
\newcommand{\V}{\ensuremath{\mathcal{V}}}
\newcommand{\E}{\ensuremath{\mathcal{E}}}

\newcommand{\revision}[1]{{#1}}
\newcommand{\etal}{~\emph{et~al.}\xspace}

\usepackage{caption}

\makeatletter
\long\def\@IEEEtitleabstractindextextbox#1{\parbox{0.922\textwidth}{#1}}
\makeatother

\begin{document}

\title{The Power of Connection: \\ Leveraging Network Analysis \\ to Advance Receivable Financing}

\author{Ilaria~Bordino, Francesco~Gullo, and Giacomo~Legnaro
\IEEEcompsocitemizethanks{
\IEEEcompsocthanksitem \mbox{\!\!\!\!\!I.~Bordino and F.~Gullo} are with UniCredit, R\&D Dept., Italy. Email:  \mbox{\!\!\!\!\!\{ibordino,gullof\}@acm.org.}
\IEEEcompsocthanksitem \mbox{\!\!\!\!\!G.~Legnaro~is~with~Prometeia,~Italy.~Email:~giacomo.legnaro@prometeia.com.}
}
}
\vspace{-5mm}


\vspace{-2mm}
\IEEEtitleabstractindextext{%
\begin{abstract}
Receivable financing is the process whereby cash is advanced to firms against receivables their customers have yet to pay: a receivable can be sold to a funder, which immediately gives the firm cash in return for a small percentage of the receivable amount as a fee.
Receivable financing has been traditionally handled in a centralized way, where every request is processed by the funder individually and independently of one another.

In this work we propose a novel, network-based approach to receivable financing, which enables customers of the same funder to autonomously pay each other as much as possible, and gives benefits to both the funder (reduced cash anticipation and exposure risk) and its customers (smaller fees and lightweight service establishment). 

Our main contributions consist in providing a principled formulation of the network-based receivable-settlement strategy, and showing how to achieve all algorithmic challenges posed by the design of this strategy.
We formulate network-based receivable financing as a novel combinatorial-optimization problem on a multigraph of receivables.
We show that the problem is \NPhard, and devise an exact branch-and-bound algorithm, as well as algorithms to efficiently find effective approximate solutions.
Our more efficient algorithms are based on cycle enumeration and selection, and exploit a theoretical characterization in terms of a knapsack problem, as well as a refining strategy that properly adds paths between cycles.
We also investigate the real-world issue of avoiding temporary violations of the problem constraints, and design methods for handling it.

An extensive experimental evaluation is performed on real receivable data. 
Results attest the good performance of our methods.
\end{abstract}


}

\maketitle
\sloppy
\IEEEdisplaynontitleabstractindextext
\IEEEpeerreviewmaketitle

\section{Introduction}
\label{sec:intro}

The term \emph{receivable} indicates a debt owed to a creditor, which the debtor has not yet paid for. 

\spara{Receivable Financing} (RF) is a service for creditors to fund cash flow by selling accounts receivables to a \emph{funder} or \emph{financing company}.
The funder anticipates (a proportion of) the receivable amount to the creditor, deducting a percentage as a fee for the service.

Receivable financing mainly exists to shorten the waiting times of receivable payment. 
These typically range from 30 to 120 days, which means that businesses face a nervous wait as they cannot effectively plan ahead without knowing when their next payment is coming in.
Cashing anticipated payment for a receivable gives a business instant access to a lump sum of capital, which significantly eases the cash-flow issues associated with receivables. 
Another pro is that funders typically manage credit control too, meaning that credit control is outsourced and 
creditors no longer need to chase up debtors for receivable payment. 

RF has traditionally been provided by banks and financial institutions.
Recently, digital platforms have emerged as well, e.g.,  \emph{BlueVine}, \emph{Fundbox}, \emph{C2FO},  \emph{MarketInvoice}.
Nevertheless, while digital funders are rapidly growing, they are still far from saturating the commercial credit market.

\begin{figure}[t]
\centering
\vspace{-15mm}
\captionsetup{justification=justified,singlelinecheck=off,font={stretch=0.8}}
\begin{minipage}[b]{0.44\linewidth}
\includegraphics[width=1.0\textwidth]{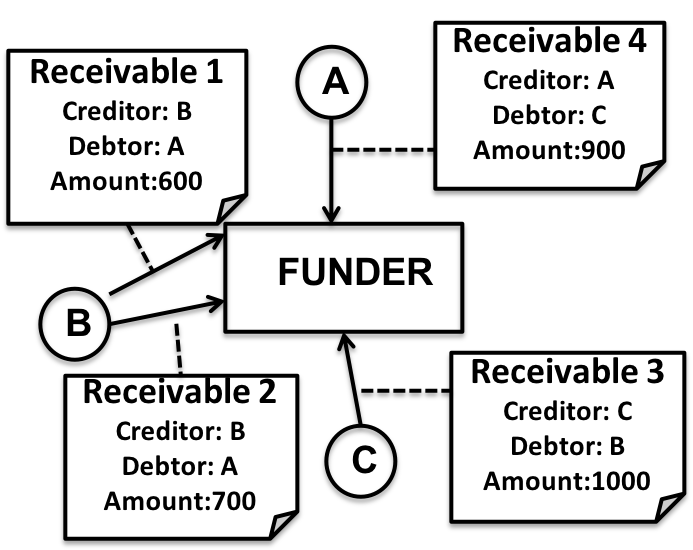} 
\end{minipage}
\hspace{3.5mm}
\begin{minipage}[b]{0.5\linewidth}
\includegraphics[width=1.0\textwidth]{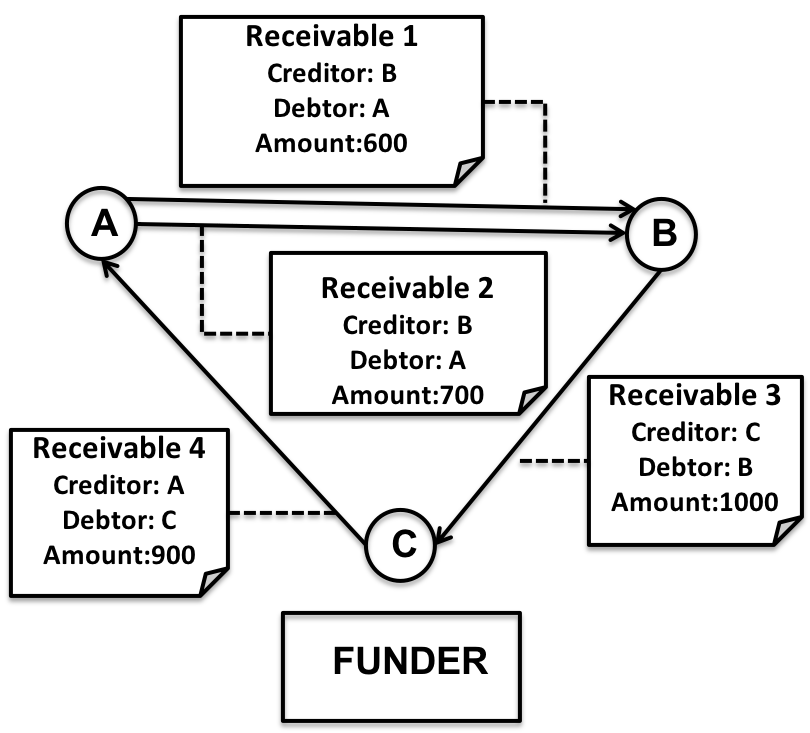} 
\end{minipage}
\vspace{-6mm}
\caption{ \label{img:network} 
{\footnotesize 
Client/server (left) vs. network-based (right) receivable financing.
In the first case, the funder handles each request individually, paying a total amount of 3200.
In the network-based scenario, customers are allowed to pay each other.
\revision{
Assume  $A$, $B$, $C$ have  $0$, $1000$, $0$ on their accounts, respectively.
Then, $B$ has sufficient money to pay Receivable~3, which in turns makes $C$ have enough money to pay Receivable~4, and so on, until all receivables have been paid. This way no money is anticipated by the funder.}
}
}
\vspace{-3mm}
\end{figure}

\spara{A novel, network-based perspective.}
Existing RF services employ an inherently \emph{client-server} perspective: the funder, just like a centralized server, receives multiple funding requests, and processes each of them individually.

It is easy to observe that this strategy has an obvious limitation, i.e., it overlooks the fact that a set of receivables for which the RF service is simultaneously requested compose a \emph{network where a customer may act as the creditor or the debtor} of different receivables. In this work we present a novel receivable-financing method, which leverages such a network perspective to enable an autonomous cash flow among customers. 
\revision{
For instance, as shown in Figure~\ref{img:network}, a network-based RF service might allow the funder to anticipate no money, while still having  RF  accomplished for all the receivables.
A real scenario is clearly more complicated than the simple one depicted in the example. 
This means that the receivables that can be autonomously paid by the customers without involving the funder are typically only a subset of all the receivable which the RF service is requested for.
The main goal in the design of  network-based RF is therefore \emph{to select the largest-amount subset of receivables for which autonomous payment is possible}.
}

Network-based RF
provides noticeable advantages to both the funder and the customers.
\revision{
In fact, the funder can now entrust to the novel network-based settlement system the settlement of the receivable financing requests advanced by the customers who accept to enroll into it. The system takes care to enable autonomous payment of these receivables to the maximum extent possible. 
This means that the funder no longer needs to anticipate cash for each request, achieving a \emph{larger} availability of \emph{liquidity}, and a \emph{lower risk} of incurring into payment delays or credit losses. 
The increased cash availability in turn allows the funder to devise proper marketing strategies to attract more customers to try the novel settlement service. 
Specifically, the funder offers the customers the benefit of a \emph{lower fee} on each incoming deliverable. 
A further advantage for the customer is the \emph{easier access} to the receivable financing service (in terms of both time and effort), which is enabled by  the employment of lighter bureaucracy and risk-assessment rules.
}

\revision{
A crucial aspect for the network-based RF service to work properly is to regulate it with to a do-ut-des mechanism, where customers accept to pay selected receivables, for which they act as a debtor, \emph{possibly earlier than they would do without the service}. 
Our enrolling customer is warned that the system does not allow non-payments, and that when a receivable for which she plays as debtor is selected for settlement, its amount is automatically transferred to the creditor. 
However, the customer is informed that she also benefits from early payments, because the system ensures that they increase the chance for a debtor of subsequently acting as a creditor.
Moreover, customers may preserve their freedom on how to handle payments, by choosing which requests to submit to the network-based system. 
}

\spara{Challenges and contributions.} 
In this work we tackle a real-world problem from a specific application domain, i.e., receivable financing, proposing a \emph{novel} method, based on a \emph{network-perspective} that -- to the best of our knowledge --
has never been employed for such a problem before. 
Effectively solving the network-based receivable-settlement problem requires non-trivial algorithmic and theoretic work. 
This paper focuses on the algorithmic challenges that arise amid the design of a such a network-based approach. 

We observe that the set of receivables that are available for settlement at any day can be naturally modeled as a \emph{multigraph}, whose nodes correspond to customers, and an arc from $u$ to $v$ represents a receivable having $u$ as a  debtor and  $v$ as a creditor.
Such a  multigraph is \emph{arc-weighted}, with weights corresponding to receivable amounts, and \emph{node-attributed}, as every node is assigned its account balance(s), as well as  a \emph{floor} and a \emph{cap}, which serve the purpose of limiting the balance(s) of a node to stay within a reasonable range.
We formulate network-based receivable settlement as a novel combinatorial-optimization problem on a multigraph of receivables, whose objective is to select a subset of arcs so as to maximize the total amount of the selected arcs, while also satisfying two constraints: ($i$) the balance of every node meets the corresponding floor-cap constraints, and ($ii$) every node within the output solution is the creditor of at least one output receivable and the debtor of at least one output receivable.
We show that the problem is \NPhard and devise both an exact algorithm and effective algorithms to find approximate solutions more efficiently.
Our more efficient algorithms exploit the fact that a cycle of the input multigraph is guaranteed to satisfy one of the two problem constraints. 
For this reason, we formulate a further combinatorial-optimization problem, which identifies a subset of the cycles of the input multigraph, so as to maximize the total amount and keep the floor-cap constraints satisfied. Such a problem is shown to be \NPhard too, and our most efficient algorithms find approximate solutions to it by exploiting a knapsack-like characterization, as well as a refining strategy that properly adds paths between the selected cycles.
Orthogonally, we focus on a practical issue that may arise while implementing a network-based RF service in real-world systems: to minimize system re-engineering, it might be required for the money transfers underlying the settled receivables to be executed one at a time and without violating the problem constraints, not even temporarily.
We devise proper strategies to handle such a real-world issue, by either redefining floor constraints, or taking a subset of the settled receivables and properly ordering them.

To summarize, the main contributions of this work are:
\vspace{-0mm} 
\begin{itemize}
\item
We devise a novel, network-based approach to receivable financing~(Section~\ref{sec:background}).
\item
We provide a principled formulation of the network-based receivable settlement strategy in terms of a novel optimization problem, i.e., \staticbalance, while also showing the  \NPhard{ness} of the problem (Section~\ref{sec:problem}).
\item
We derive theoretical conditions to bound the objective-function value of a set of problem solutions, and employ them to design an exact  \emph{branch-and-bound} algorithm (Section~\ref{sec:bbalg}).
\item
We define a further optimization problem, i.e., optimal selection of a subset of cycles, and characterize it in terms of \NPhard{ness} and as a \knapsack problem.
Based on such an optimal-cycle-selection problem, we devise an efficient algorithm for the original \staticbalance problem (Section~\ref{section:approxalg}).
\item
We also present a method to improve upon the cycle-selection-based algorithm by suitably adding paths between the selected cycles~(Section~\ref{algo:paths}).
This algorithm and the branch-and-bound one are also combined into a  further \emph{hybrid} algorithm~(Section~\ref{algo:hybrid}).
\item
We handle the real-world scenario where temporary constraint violations are not allowed~(Section~\ref{algo:negbal}).
\item
We present an extensive evaluation on a real dataset of receivables. Results demonstrate the effectiveness of the proposed methods in practice (Section~\ref{sec:experiments}).
\end{itemize}

\noindent
In Section~\ref{sec:implementation} we report implementation details, while
Section~\ref{sec:related} offers related work, and Section~\ref{sec:conclusion} draws~conclusions. 

This is a complete technical report that includes contents from~\cite{bordino18network} and \cite{bordino20advancing}.

\vspace{-0mm}
\section{Preliminaries}
\label{sec:background}


In this work we consider the following scenario.
We focus on the customer basis $\mathcal{U}$ of a single funder (we assume the funder has no visibility on the customers of other funders).
Customers in $\mathcal{U}$ submit RF requests to the funder.
We denote by $\mathcal{R}$ the set of receivables submitted to the funder by its customers.
The goal of the funder is to select a subset of $\mathcal{R}$ to be settled, by employing a \emph{network-based} strategy, i.e., making the involved customers pay each other autonomously.


\spara{Receivables.}
Attributes of a receivable $R \in \mathcal{R}$ include:
\begin{itemize}
\item
$\amount(R) \in \mathds{R}$: amount of the receivable;
\item
$\creditor(R) \in \mathcal{U}$: payee of the receivable;
\item
$\debtor(R) \in \mathcal{U}$: payer of the receivable;
\item
$\insertdate(R)$: date the receivable entered the system;
\item
$\duedate(R)$: date on which the payment falls due;
\item
$\lifetime(R) \in \mathbb{N}$: the maximum number of days the network-based RF service is allowed to try to settle the receivable.
\end{itemize}
A receivable $R$ is said \emph{active} for $\creditor(R)$, and \emph{passive} for $\debtor(R)$.

\spara{Customers.}
\spara{Customers.}
Customers are assigned a dedicated account by the funder, which is used to pay passive receivables or get paid for active receivables. 
Moreover, customers may perform \emph{deposits}/\emph{withdrawals} on/from their account.
Such operations trigger \emph{external} money flows, which do not derive from receivable settlement.
The desideratum to keep the system in a collaborative equilibrium is that, if a customer withdraws money from her account, this operation should not increase the customer's marginal availability to cash in more receivables through the RF service. Conversely, if a customer deposits money on her account, this operation should obviously increase her ability to pay more receivables.
To take into account these principles, we keep track of two different balances in the account of customer $u$, and require such balances to be limited by a \emph{floor} and a \emph{cap} (which are set on a customer basis during sign up).
As a result, every customer $u \in \mathcal{U}$ is assigned the attributes:
\begin{itemize}
\item
$\balr(u) \in \mathds{R}$: \emph{receivable balance} of $u$'s account, i.e., the sum of all receivables $u$ has got paid minus the sum of all receivables $u$ has paid through the RF service over the whole $u$'s lifetime;
\item
$\bala(u)  \in \mathds{R}$: \emph{actual balance} of $u$'s account, corresponding to the receivable balance $\balr(u)$, increased by money from deposit operations and decreased by withdrawals;
\item
$\ub(u) \in \mathds{R}$: upper bound on the receivable balance of $u$'s account; requiring $\balr(u) \leq \ub(u)$ at any time avoids unbalanced situations where a customer utilizes the service only to get money without paying passive receivables;
\item
$\lb(u) \in \mathds{R}$: lower bound on the actual balance of $u$'s account; typically, $\lb(u) = 0$, but in some cases negative values are allowed, meaning that some overdraft is tolerated.
\end{itemize}

\begin{table}[t!]
\centering
\vspace{5mm}
\caption{\footnotesize\label{table:notation}\revision{List of frequent symbols.}}
\vspace{-0cm}
\scriptsize
\revision{
\begin{tabular}{@{}c@{ }|@{ }l@{}}
\emph{symbol}&\emph{definition}\\ \hline
$\G \!=\! (\V, \!\E, \!w)$ & input \bmultigraph \\
$\balr(u)$ & receivable balance of $u$'s account ($u \in \V$)\\
$\bala(u)$ & actual balance of $u$'s account ($u \in \V$)\\
$\ub(u)$ & upper bound on $\balr(u)$ ($u \in \V$)\\
$\lb(u)$ & lower bound on $\bala(u)$ ($u \in \V$)\\
$\cycles$ & set of cycles in $\G$\\
$C \in \cycles$ & a cycle in $\cycles$\\
$\paths_{ij}$ & paths from a node in cycle $C_i$ to a node in cycle $C_j$\\
$\paths_{max}$ & max size of a path set in $\{\paths_{ij}\}_{i,j}$ \\
$L$; \ $L_p$ & max length of a cycle in $\cycles$; \ max length of a path in $\paths_{ij}$\\
$K;$  $K_p$ & size of $\cycles$'s subset; \ size of $\paths_{ij}$'s subset (beam-search algorithms)\\
$\mathbf{CC}$ & weakly connected components of $\G$\\
$H$ & max size of a $\G$'s conn. component to run the exact algorithm on\\
\end{tabular}
}
\vspace{0mm}
\end{table}

\spara{Network-based RF in action.}
The execution flow of our service is as follows. 
A receivable $R$ is submitted by $creditor(R)$ to the netwok-based RF service.
While submitting $R$, the creditor also sets $\lifetime(R)$, i.e., the maximum number of days $R$ can stay in the system: if $R$ has not been settled during that period, the creditor gets it back (and she may require to sell it to different financing services).

Once $R$ has been added to the system, $debtor(R)$ is asked to confirm if she agrees with paying $R$ anytime between $\insertdate(R)$ and $\duedate(R)$.
If $debtor(R)$ gives her confirmation, it means that she accepts to pay $R$ possibly before its \duedate.
This is a crucial aspect in the design of network-based RF.
%
%
\revision{
In this regard, a specific mechanism is employed to maintain the desired equilibrium where customers autonomously pay each other as far as possible:  the debtor must accept to pay a receivable before its \duedate to \emph{gain operability within the service, so as to get her (future) active receivables settled more easily}.
Indeed, according to the constraint $\balr(u) \leq \ub(u)$, 
the more the receivables paid by $u$ through the network-based RF service, the further $\balr(u)$ remains from $\ub(u)$ and the higher the chance for $u$ to have her active receivables paid. 
}

After confirmation by $\debtor(R)$, $R$ becomes part of the set $\mathcal{R}$ of receivables that the system will attempt to settle (according to the method(s) presented in Sections~\ref{sec:problem}-\ref{sec:algo}).
If the system is not able to settle $R$ during the period $[\insertdate(R), \min\{\insertdate(R)+\lifetime(R), \duedate(R)\}]$, $creditor(R)$ gets the receivable back.
Otherwise, $\amount(R)$ is withdrawn from the $debtor(R)$'s account and put to the $creditor(R)$'s account. 
Without loss of generality, we assume that the settlement fee of receivable $R$ is paid by $\creditor(R)$ to the funder of the RF service through a different channel.
\revision{
As better explained in Section~\ref{sec:problem}, 
a receivable $R$ can be selected for settlement only if it complies with the economic conditions of $debtor(R)$'s (and $creditor(R)$'s) account  (and some other global constraints are satisfied, see Problem~\ref{problem:balancestatic}).
This prevents the  system from being affected by insolvencies.
}

\revision{
\begin{figure}[t]
\vspace{2mm}
\captionsetup{justification=justified,singlelinecheck=off,font={stretch=0.85}}
{
\begin{center}
\begin{tabular}{cc}
\!\!\!\!\!\!\!\!\!\!\begin{tikzpicture}[scale=0.19]
	\node[circle, draw, thick, scale=0.7] (A) at (0,12) {\textsf{A}};
	\node[scale=0.7, rotate=0] (Alabel) at (0,14.8) {\begin{tabular}{c} 1000 \\ $[$-1000, 3000$]$ \end{tabular}};
	\node[circle, draw, thick, scale=0.7] (B) at (16,12) {B};
	\node[scale=0.7, rotate=0] (Blabel) at (16,14.8) {\begin{tabular}{c} 0 \\ $[$-500, 4000$]$ \end{tabular}};	
	\node[circle, draw, thick, scale=0.7] (C) at (4,6) {C};
	\node[scale=0.7, rotate=0] (Clabel) at (6,4) {\begin{tabular}{c} 100 \\ $[$0, 1500$]$ \end{tabular}};	
	\node[circle, draw, thick, scale=0.7] (D) at (16,0) {D};
	\node[scale=0.7, rotate=0] (Dlabel) at (16,-2.8) {\begin{tabular}{c} 100 \\ $[$-500, 3000$]$ \end{tabular}};	
	\node[circle, draw, thick, scale=0.7] (E) at (0,0) {E};
	\node[scale=0.7, rotate=0] (Elabel) at (0,-2.8) {\begin{tabular}{c} 0 \\ $[$-200, 600$]$ \end{tabular}};	
	
	\node[scale=0.7,rotate=323] (DA) at (12,4) {2600};
	\node[scale=0.7] (DE) at (8,-1) {1600};
	\node[scale=0.7] (AB1) at (8,14.5) {700};
	\node[scale=0.7] (AB2) at (8,11.3) {600};
	\node[scale=0.7,rotate=90] (EA) at (-1,6) {1000};
	\node[scale=0.7,rotate=90] (BD1) at (14,6) {700};
	\node[scale=0.7,rotate=90] (BD1) at (16.5,6) {1000};
	\node[scale=0.7,rotate=20] (CB) at (11,7.7) {1100};
	\node[scale=0.7,rotate=55] (EC) at (1.2,3.2) {900};
	
	\draw[->, thick, >=stealth, bend left=18] (A) edge (B);
	\draw[->, thick, >=stealth, bend right=18] (A) edge (B);
	\draw[->, thick, >=stealth, bend left=18] (B) edge (D);
	\draw[->, thick, >=stealth, bend right=18] (B) edge (D);
	\draw[->, thick, >=stealth] (D) edge (E);
	\draw[->, thick, >=stealth] (E) edge (A);
	\draw[->, thick, >=stealth, dashed] (E) edge (C);
	\draw[->, thick, >=stealth, dashed, bend right=10] (C) edge (B);
	\draw[->, thick, >=stealth, dashed] (D) edge (A);
\end{tikzpicture}
&
\!\!\!\!\!\!\begin{tikzpicture}[scale=0.19]
	\node[circle, draw, thick, scale=0.7] (A) at (0,12) {\textsf{A}};
	\node[scale=0.7, rotate=0] (Alabel) at (0,14.8) {\begin{tabular}{c} 700 \\ $[$-1000, 3000$]$ \end{tabular}};
	\node[circle, draw, thick, scale=0.7] (B) at (16,12) {B};
	\node[scale=0.7, rotate=0] (Blabel) at (16,14.8) {\begin{tabular}{c} -400 \\ $[$-500, 4000$]$ \end{tabular}};	
	\node[circle, draw, thick, scale=0.7] (C) at (4,6) {C};
	\node[scale=0.7, rotate=0] (Clabel) at (6,4) {\begin{tabular}{c} 100 \\ $[$0, 1500$]$ \end{tabular}};	
	\node[circle, draw, thick, scale=0.7] (D) at (16,0) {D};
	\node[scale=0.7, rotate=0] (Dlabel) at (16,-2.8) {\begin{tabular}{c} 200 \\ $[$-500, 3000$]$ \end{tabular}};	
	\node[circle, draw, thick, scale=0.7] (E) at (0,0) {E};
	\node[scale=0.7, rotate=0] (Elabel) at (0,-2.8) {\begin{tabular}{c} 600 \\ $[$-200, 600$]$ \end{tabular}};	
	
	\node[scale=0.7,rotate=323] (DA) at (12,4) {2600};
	\node[scale=0.7,rotate=20] (CB) at (11,7.7) {1100};
	\node[scale=0.7,rotate=55] (EC) at (1.2,3.2) {900};
	\node[scale=0.7,rotate=90] (AE) at (-1,6) {1000};
	\node[scale=0.7,rotate=305] (CA) at (2,7.7) {50};
	
	\draw[->, thick, >=stealth, dashed] (E) edge (C);
	\draw[->, thick, >=stealth, dashed, bend right=10] (C) edge (B);
	\draw[->, thick, >=stealth, dashed] (D) edge (A);
	\draw[->, thick, >=stealth, densely dotted] (A) edge (E);
	\draw[->, thick, >=stealth, densely dotted] (C) edge (A);
\end{tikzpicture}\vspace{-1mm}\\
\!\!\!\!\!\!\!\!\!\!\revision{(a) {\small Day 1}}  & \!\!\!\!\!\!\revision{(b) {\small Day 2}}
\end{tabular}
\end{center}
}
\vspace{-2mm}\caption{ \label{img:example} 
\revision{
\footnotesize 
An instance of \staticbalance (Problem~\ref{problem:balancestatic}). 
Nodes are assigned balances (in this example \balr = \bala), and $floor$-\ub ranges (square brackets). 
Arcs are labeled with the amount of the corresponding receivables.
Day~1: the arcs depicted with full lines form the optimal solution. No other arcs can be selected. In fact, adding $(\mathrm{E},\mathrm{C})$ and $(\mathrm{C},\mathrm{B})$ to the solution would violate $\mathrm{C}$'s and $\mathrm{E}$'s $floor$ constraint, while adding $(\mathrm{D},\mathrm{A})$ would violate $\mathrm{D}$'s $floor$ constraint.
Day~2: receivables present in the system corresponds to the ones not settled at Day~1, along with the two new ones depicted with dotted lines (i.e., $(\mathrm{A},\mathrm{E})$ and $(\mathrm{C},\mathrm{A})$). The optimal solution at Day~2 is an empty set. In fact, the only way to have Problem~~\ref{problem:balancestatic}'s Constraint~(2) satisfied would be selecting $(\mathrm{A},\mathrm{E})$, $(\mathrm{E},\mathrm{C})$, and $(\mathrm{C},\mathrm{A})$ altogether. However, this is not possible as it would not comply with $\mathrm{E}$'s \ub constraint.
}
}
\vspace{-0mm}
\end{figure}
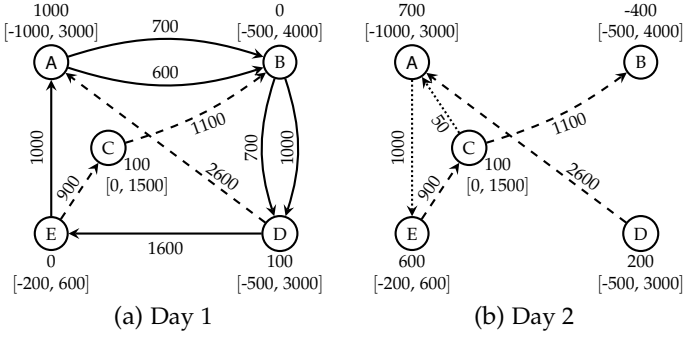
}

\section{Problem Definition}
\label{sec:problem}
\vspace{3mm}


A network-based RF service requires the design of a proper strategy to select a subset of receivables to be settled.
Here we assume receivable settlement works on a daily basis, running \emph{offline at the end of a working day} $t$, and taking as input receivables that are \emph{valid} at time $t$, i.e., $\mathcal{R}(t) = \{R \in \mathcal{R} \mid t \in [\insertdate(R),$ $\min\{\insertdate(R)$ $ + \lifetime(R), \duedate(R)\}]\}$. 
This scenario induces a \emph{multigraph},  where arcs correspond to receivables, and nodes to customers.
This multigraph, termed \mbox{\emph{\bmultigraph}}, is \emph{directed}, \emph{weighted}, and \emph{node-attributed}:

\vspace{0mm}
\begin{definition}[\bmultigraph]\label{def:balancemultigraph}
Given a set $\mathcal{R}(t)$ of receivables active at time $t$, the \emph{\bmultigraph} induced by  $\mathcal{R}(t)$ is a triple $\G = (\V, \E, w)$, where $\V$ is a set of \emph{nodes}, $\E$ is a \emph{multiset}  of \emph{ordered} pairs of nodes, i.e., \emph{arcs}, and $w: \E \rightarrow \mathds{R}^+$ is a function assigning (positive real) weights to arcs.
Each arc $(u,v) \in \E$ models the case ``$u$ pays $v$'', i.e., it corresponds to a receivable $R$ where $u = \debtor(R)$, $v = \creditor(R)$, and $w(u,v) = \amount(R)$.
Each node $v \in \V$ is assigned attributes \mbox{$\balr(u)$, $\bala(u)$, $\ub(u)$, and $\lb(u)$}.
\end{definition}
\vspace{2mm}

The objective in our network-based settlement is to select a set of receivables, i.e., arcs of the \bmultigraph $\G$, so as to \emph{maximize the total amount}.
This is a desideratum for both the funder and its customers.
A larger amount of settled receivables brings more profit to the founder, and also
larger savings for the customers, who -- in the absence of this service -- would be forced to pay for more expensive (traditional) alternatives.
The identified receivables should satisfy two constraints for every customer $u$ spanned by them: (1) the resulting $\balr(u)$ and $\bala(u)$ should be within $\ub(u)$ and $\lb(u)$ (i.e., $\balr(u) \leq \ub(u)$, $\bala(u) \geq \lb(u)$), and (2) $u$ should be the payer of at least one selected receivable and the payee of at least another selected receivable. 
Constraint (2) arises due to a specific marketing choice, i.e., preventing a customer from only paying receivables in a given day. 
This is based on the idea that showing clients that any day they pay a receivable, they also receive at least one payment as creditors, may crucially help increase their appreciation and engagement with the service. 
Moreover, preventing a customer from only paying receivables serves the aim of ensuring the aforementioned \emph{do-ut-des} principle.

The above desiderata are formalized into the following combinatorial-optimization problem, while Figure \ref{img:example} depicts a (toy) problem instance.

\vspace{1mm}
\begin{problem}[\staticbalance]\label{problem:balancestatic}
Given an \bmultigraph $\mathcal{G} \!=\! (\V, \E, w)$, find
\vspace{-0mm}
\begin{eqnarray}
\E^* & = & \arg\max_{\hat{\E} \subseteq \E} \ \ {\displaystyle\sum_{e \in \hat{\E}}} w(e) \qquad \mbox{subject to}\nonumber\\\
& & \!\!\!\!\!\!\!\!\!\!\!\!\!\!\!\!\!\!\!\!\!\!\!\!\!\!\!\!\!\!\!\! \Big (\!\!\!\!\!\!\sum_{(v,u) \in \hat{\E}} \!\!\!\!\!w(v,u) \! - \!\!\!\!\!\!\!\sum_{(u,v) \in \hat{\E}} \!\!\!\!\!w(u,v)\! \Big) \!\!\in\!\! \left [\lb(u) \!-\! \bala(u), \ub(u) \!-\! \balr(u) \right ]\!, \quad \ \label{eq:prob1constraint1}\\
& & \!\!\!\!\!\!\!\!\!\!\!\!\!\!\!\!\!\!\!\!\!\!\!\!\!\!\!\!\!\!\!\! |\{(u,v) \mid (u,v) \in \hat{\E}\}| \geq 1\!, \label{eq:prob1constraint2}  \mbox{ and } |\{(v,u) \mid (v,u) \in \hat{\E}\}| \geq 1\!, \label{eq:prob1constraint3}
\end{eqnarray}
$\forall u \in \V(\hat{\E}) = \{u \in \V \mid (u,v) \in \hat{\E} \vee (v,u) \in \hat{\E}\}$.
\end{problem}


\begin{theorem}\label{th:hardness}
Problem~\ref{problem:balancestatic} is \NPhard.
\end{theorem}
\begin{IEEEproof}
We reduce from the well-known \NPhard \subsetsum problem~\cite{Kellerer04}:  given a set $S$ of positive real numbers and a further real number $B > \min\{x \mid x \in S\}$, find a subset $S^* \subseteq S$ such that the sum of the numbers in $S^*$ is maximum and no more than $B$.
%
Given an instance $\langle S = \{x_1, \ldots, x_k\}, B \rangle$ of \subsetsum, we construct a \staticbalance instance composed of a multigraph $\G = (\V, \E, w)$ having two nodes, i.e., $u$ and $v$, one arc from $v$ to $u$ with weight equal to some positive real number $\epsilon < \min\{x \mid x \in S\}$, 
and as many additional arcs from $u$ to $v$ as the numbers in $S$, with the weight on each arc $e_i$ from $u$ to $v$ being equal to the corresponding number $x_i \in S$.
Moreover, we let $u$ and $v$ have the following attributes: $\balr(u) = \bala(u) = 0$, $\ub(u) = \epsilon$, $\lb(u)=-\sum_{(u,v) \in \E} w(u,v)$, 
$\balr(v) = \bala(v) = \epsilon$, $\ub(v) = B$, and $\lb(v)=-\epsilon$.
The optimal solution $\E^*$ for the \staticbalance instance $\G$ possesses the following features:
\vspace{-0mm}
\begin{smallitemize}
\item
$\E^* \neq \emptyset$, as there exists at least one non-empty feasible solution whose objective function value is larger than the empty solution, e.g., the solution $\{e_{min}, (v,u)\}$, where $e_{min}$ $=\arg\min_{i \in [1..k]} w(e_i)$ (as $w(v,u) = \epsilon < \min_{i \in [1..k]} w(e_i)$ by construction).
\item
Arc $(v,u)$ will necessary be part of $\E^*$, otherwise Constraint~(\ref{eq:prob1constraint2}) in Problem~\ref{problem:balancestatic} would be violated.
\item
Apart from $(v,u)$, $\E^*$ will contain all those arcs $(u,v)$ that ($i$) fulfill Constraint~(\ref{eq:prob1constraint1}) in Problem~\ref{problem:balancestatic}, and ($ii$) the sum of their weights is maximized.
The constraints to be satisfied on node $u$ are:
\vspace{-1mm}
$$
\vspace{-2mm}
\ \ \ \textstyle w(v,u) - \sum_{(u,v) \in \E^*} w(u,v) \in [\lb(u) - \bala(u), \ub(u) - \balr(u)],
$$
that is $\epsilon - \sum_{(u,v) \in \E^*} w(u,v) \in \left [-\sum_{(u,v) \in \E} w(u,v), \epsilon \right]$,
which is always satisfied, as $\epsilon \in (0, \min_{(u,v) \in \E} w(u,v))$.

\noindent The constraints to be satisfied on node $v$ are instead:
\vspace{-1mm}
$$
\!\!\!\!\!\!\!\!\!\!\!\!\textstyle\quad\qquad\sum_{(u,v) \in \E^*} w(u,v)  - w(v,u)  \in  [\lb(v) - \bala(v), \ub(v) - \balr(v)],
$$
that is $\sum_{(u,v) \in \E^*} w(u,v) - \epsilon \in [-2\epsilon, B - \epsilon]$.
The constraint $\sum_{(u,v) \in \E^*}$ $\!w(u,v) \!-\! \epsilon \!\geq\! -2\epsilon$ is always satisfied, as all $w(u,v)$ and $\epsilon$ are $\geq 0$.
The constraint $\sum_{(u,v) \in \E^*} w(u,v) - \epsilon \leq B - \epsilon$ corresponds to $\sum_{(u,v) \in \E^*} w(u,v) \leq B$, i.e., the \subsetsum constraint.
\end{smallitemize} 

\noindent
As a result, the optimal $\E^*$  of the constructed \staticbalance instance contains arcs whose sum of weights is maximum and $\leq B$, which corresponds to the optimal solution to the original \subsetsum instance.
The theorem follows.
\end{IEEEproof}

\section{Algorithms}
\label{sec:algo}

\subsection{Exact algorithm}
\label{sec:bbalg}
The first proposed algorithm for \staticbalance  is a  \emph{branch-and-bound} exact algorithm, dubbed \bbalg.

\begin{figure}[t!]
\vspace{-0mm}
\centering
\captionsetup{justification=justified,singlelinecheck=off,font={stretch=0.7}}
\includegraphics[width=.5\textwidth]{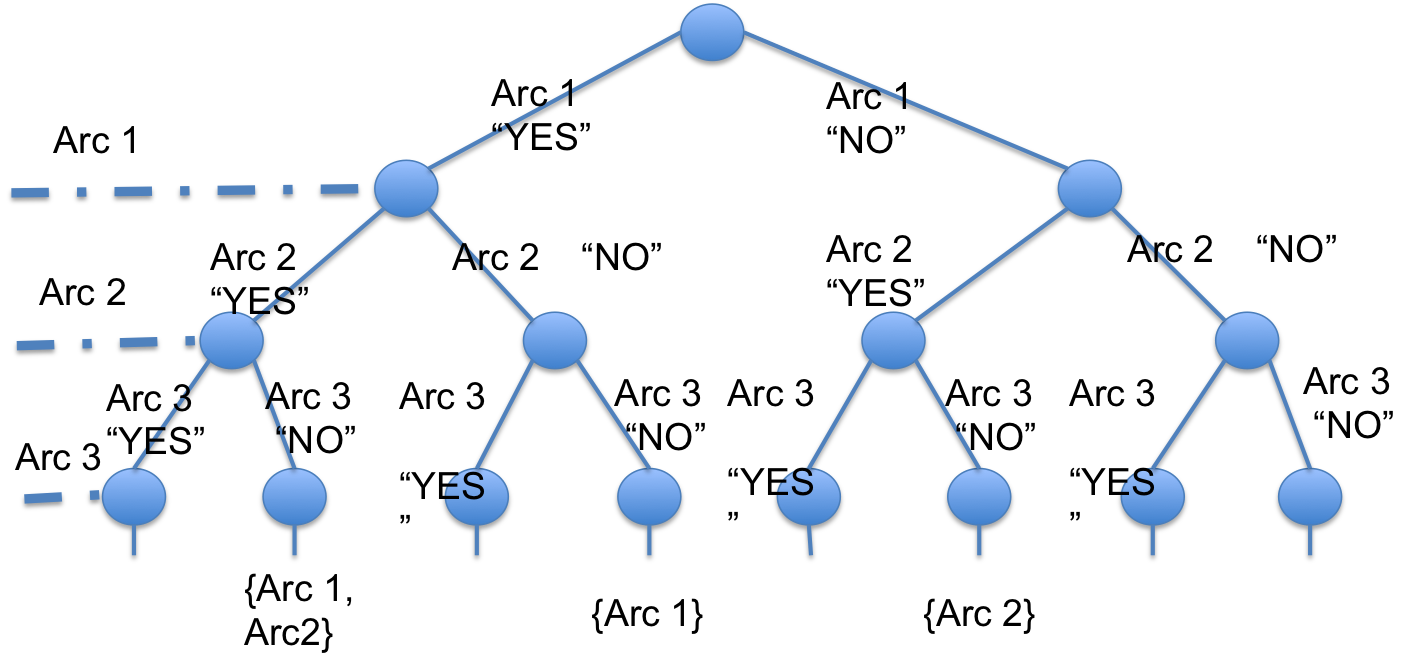}
\vspace{-6mm}
\caption{\footnotesize Tree-like representation of the \staticbalance search space considered in the \bbalg algorithm.\label{fig:bbtree}}
\vspace{5mm}
\end{figure}

\spara{Search space.}
Given an \bmultigraph $\G = (\V, \E, w)$, the search space of \staticbalance corresponds to the set $2^{\E}$ of all possible (multi)subsets of arcs.
The \bbalg algorithm represents this search space as a \emph{binary tree} $\mathcal{T}$ with $|\E|\!+\!1$ levels, where each level (except for the root one) logically represents an arc $e \in \E$ for which a decision has to be taken, i.e., include the arc or not in the output solution (Figure~\ref{fig:bbtree}).
Correspondence between arcs and levels comes from some ordering on the arcs (e.g., by non-increasing weight, as in Section~\ref{algo:hybrid}).
A path from the root to a leaf represents a complete individual solution $\hat{\E} \in 2^{\E}$ (where a decision has been taken for all arcs). 
A non-leaf tree-node\footnote{We use the term ``tree-node'' to refer to the nodes of the tree-like search space (to distinguish them from the nodes of the input \bmultigraph).} represents a set of solutions: those corresponding to all possible decisions for the arcs not in the path from the root to that non-leaf node.

\spara{Search-space exploration.}
The \bbalg algorithm explores the tree-like search space (according to some visiting strategy, e.g., \textsc{bfs} or \textsc{dfs}) by exploiting a \emph{lower bound} and an \emph{upper bound} on the set of all solutions identified by a non-leaf tree-node it has visited, and keeping track of the largest lower-bound among all the ones computed so far.
Whenever the upper bound of a tree-node is smaller than the largest so-far lower bound, that node and the whole subtree rooted in it can safely be discarded.
The visit stops when all tree-nodes have been visited or pruned, and the optimal solution $\E^*$ is selected among all survivor leaves, specifically as the one satisfying all the constraints of \staticbalance and exhibiting the maximum objective-function value.
Note that, during the visit of the search space, intermediate solutions whose partial decisions make them (temporarily) infeasible \emph{are not discarded}, because such solutions may become feasible later on.
%
The general scheme of \bbalg is outlined as Algorithm~\ref{alg:staticbalanceBB}.
A specific visiting strategy of the search space (e.g., \textsc{bfs} or \textsc{dfs}) can be implemented by properly defining the way of choosing the next tree-node to be processed (Line~4).
A crucial point in \bbalg is the definition of lower bound and upper bound. We discuss this in the remainder of the subsection.

\begin{algorithm}[h!]
\caption{\normalfont \bbalg}
\label{alg:staticbalanceBB}
\begin{algorithmic}[1]
\small
\REQUIRE An \bmultigraph $\G = (\V, \E, w)$
\ENSURE A multiset $\E^* \subseteq \E$
\vspace{1mm}
\STATE $\mathcal{T} := \mbox{tree-like representation of } 2^{\E}$
\STATE $\mathcal{X} \gets \{\mbox{root of } \mathcal{T}\}$,  \ \ $LB_{max} \gets 0$
\WHILE{$\mathcal{X} \mbox{ contains some non-leaf tree-nodes}$}
\STATE $X \gets \mbox{extract (and remove) a non-leaf tree-node from } \mathcal{X}$
\STATE $UB_X \gets \mbox{upper bound on the solutions spanned by $X$}$\hfill\COMMENT{Algortihm~\ref{alg:mincostflow}}
\IF{$UB_X \geq LB_{max}$}
\STATE $LB_X \gets  \mbox{lower bound on the solutions spanned by $X$}$\hfill\COMMENT{Algorithm~\ref{alg:johnson}}
\IF{$LB_X = UB_X$}
\STATE $\E^* \gets arcs(X)$ and stop the algorithm
\ENDIF 
\STATE $LB_{max} \gets \max\{LB_{max}, LB_X\}$
\STATE add all $X$'s children to $\mathcal{X}$
\ENDIF
\ENDWHILE
\STATE $\mathcal{L} \gets \{ \mbox{leaf } X \!\in\! \mathcal{X} \!\mid\! \mbox{$arcs(X)$ meet Problem~\ref{problem:balancestatic}'s constraints}\}$
\STATE $\E^* \gets \argmax_{arcs(X): X \in \mathcal{L}} \sum_{e \in arcs(X)} w(e)$
\end{algorithmic}
\end{algorithm}

\spara{Lower bound.}
For a tree-node $X$ at level $i$ of $\mathcal{T}$, let $\E_X$ denote the arcs for which a decision has been taken, i.e., arcs corresponding to all levels from the root to level $i$.
Let also $\E_X$ be partitioned into $\E^+_X$ and $\E^-_X$, i.e., arcs included and not included in the current (partial) solution.
A lower bound on the solutions spanned by (the subtree rooted in) $X$ can be defined by computing \emph{any feasible solution} $\hat{\E}$ to \staticbalance, subject to the additional constraint of containing all arcs in $\E^+_X$ and no arcs in $\E^-_X$.

To compute such a feasible solution, we aim at finding the set $\cycles$ of \emph{cycles} of the multigraph induced by the arc set  $\E \setminus \E^-_X$, and greedily selects cycles based on their amount, as long as they meet Constraint~(\ref{eq:prob1constraint1}) of \staticbalance.
The intuition behind this strategy is twofold.
First, a solution composed of a set of cycles always satisfies the other constraint of the problem, as every node of a cycle has at least one incoming arc and one outgoing arc.
Moreover, cycle enumeration is a well-known problem, for which a variety of algorithms exists.
As a trade-off between effectiveness, efficiency and simplicity, we employ a variant of the classic Johnson's algorithm~\cite{Johnson75}, in particular the version that works on multigraphs~\cite{Hawick08}.

More in detail, the algorithm at hand is dubbed \johnson 
and outlined as Algorithm~\ref{alg:johnson}.
To guarantee the inclusion of the arcs $\E_X^+$,
the greedy cycle-selection step (Lines~8--11) is preceded by a \emph{covering} phase (Lines~3--7), 
whose goal is to find a first subset  $\hat{\cycles} \subseteq \cycles$ of cycles that ($i$) cover all arcs in $\E_X^+$, i.e., $\E_X^+ \subseteq \bigcup_{C \in \hat{\cycles}} C$, ($ii$) maximize the total amount of the cycle set, i.e., $\sum_{e \in \bigcup_{C \in \hat{\cycles}} C} w(e)$, and ($iii$) remain feasible for \staticbalance.
This corresponds to a variant of the well-known \wsetcover problem, where $\E_X^+$ represents the universe of elements, while the cycles in $\cycles$ represent the covering sets.
The covering phase of  \johnson is therefore tackled by adapting the classic greedy $(1 + \log|\E_X^+|)$-approximation algorithm~\cite{chvatal1979greedy} for \wsetcover, which iteratively selects the set minimizing the ratio between the set cost and the number of uncovered elements within that set, until all elements have been covered.
Our adaptation consists in ($i$) defining the cost of a set as the inverse of the amount of the corresponding cycle (computed by discarding arcs already part of the output solution), and ($ii$) checking whether the \staticbalance constraints are satisfied while selecting a cycle.
In the event that not all arcs in $\E_X^+$ have been covered, the algorithm returns an empty set (and the lower bound used in \bbalg is set to zero).

\begin{algorithm}[t]
\caption{\normalfont \johnson}
\label{alg:johnson}
\begin{algorithmic}[1]
\small
\REQUIRE An \bmultigraph $\G = (\V, \E, w)$, two multisets $\E_X^+ \subseteq \E$, $\E_X^- \subseteq \E$
\ENSURE A multiset $\hat{\E} \subseteq \E \setminus \E^-_X$
\vspace{1mm}
\STATE $\cycles \gets \mbox{cycles of multigraph } \G^- = (\V, \E \setminus \E_X^-, w)$\hfill\COMMENT{cf.~\cite{Hawick08}}
\STATE $\hat{\E} \gets \emptyset$, \ \ $\hat{\cycles} \gets \emptyset$
\WHILE{$\cycles \neq \emptyset \wedge \E^+_X \nsubseteq \hat{\E}$}
\STATE $\cycles \gets  \{\mbox{$C \in \cycles \mid \hat{\E} \cup C$ meets Constraint~(\ref{eq:prob1constraint1}) of Problem~\ref{problem:balancestatic}}\}$
\STATE $C \gets \mbox{cycle in $\cycles$ minimizing $[|C \!\cap \!(\E^+_X \setminus \hat{\E})|\!\sum_{e \in C \setminus \hat{\E}} w(e)]^{-1}$}$
\STATE $\hat{\cycles} \gets \hat{\cycles} \cup \{C\}$, \ \ $\cycles \gets \cycles \setminus \{C\}$, \ \ $\hat{\E} \gets \hat{\E} \cup C$
\ENDWHILE
\WHILE{$\cycles \neq \emptyset$}
\STATE $\cycles \gets  \{\mbox{$C \in \cycles \mid \hat{\E} \cup C$ meets Constraint~(\ref{eq:prob1constraint1}) of Problem~\ref{problem:balancestatic}}\}$
\STATE $C \gets \mbox{cycle in $\cycles$ maximizing $\sum_{e \in C \setminus \hat{\E}} w(e)$}$
\STATE $\cycles \gets \cycles \setminus \{C\}$, \ \ $\hat{\E} \gets \hat{\E} \cup C$
\ENDWHILE
\IF{$\E^+_X \nsubseteq \hat{\E}$}
\STATE $\hat{\E} \gets \emptyset$
\ENDIF
\end{algorithmic}
\end{algorithm}

\smallskip
\underline{\em Time complexity}:
The running time of  \johnson is dominated by cycle enumeration (Line~2), as the number of cycles in a (multi)graph can be exponential. 
In our context this is however not blocking: as it is unlikely that the problem constraints are satisfied on long cycles, we employ a simple yet effective workaround of detecting cycles up to a certain size $\maxcyclelength$. 
%
The complexity of the remaining steps is as follows.
The covering phase (Lines~3--6) can be implemented by using a priority queue with logarithmic-time insertion/extraction.
It comprises: 
($i$) computing set-cover score and checking the problem constraints for all cycles ($\bigO(\maxcyclelength~\!|\cycles|)$ time); 
($ii$) adding/extracting cycles to/from the queue ($\bigO(|\cycles|\log|\cycles|)$ time); 
($iii$) once a cycle $C$ has been processed, updating the score of all cycles sharing some edges with $C$ ($\bigO(\maxcyclelength~\!|\cycles|\log|\cycles|)$ time, as each cycle is updated at most $\maxcyclelength$ times, and, for every time, it should be removed from the queue and re-added with updated score).
Hence, the covering phase takes $\bigO(\maxcyclelength~\!|\cycles|\log|\cycles|)$ time.
In greedy selection (Lines~7--11) cycles are processed one by one in non-increasing amount order, and added to the solution after checking (in $\bigO(\maxcyclelength)$ time) the problem constraints. This yields a $\bigO(|\cycles|~\!(\log|\cycles| + \maxcyclelength))$ time.

\spara{Upper bound.}
The \bbalg upper bound lies on a relaxation of \staticbalance where Constraint~(\ref{eq:prob1constraint2}) is discarded and arcs are allowed to be selected \emph{fractionally}:
\begin{problem}[\rstaticbalance]\label{problem:relaxedbalancestatic}
Given an \bmultigraph $\mathcal{G} = (\V, \E, w)$, find $\mbox{ } \{x_e \in [0,1]\}_{e \in \E}$ so as to
$$
\begin{array}{rl}
\mbox{Maximize} & \sum_{e \in \E} x_ew(e) \\
\mbox{subject to} & \big (\sum_{e=(v,u) \in \E} x_ew(e) - \sum_{e=(u,v) \in \E} x_ew(e) \big) \\
&  \in \left [\lb(u) \!-\! \bala(u), \ub(u) \!-\! \balr(u) \right ], \ \  \forall u \in \V
\end{array}
$$
\end{problem}
The desired upper bound relies on an interesting characterization of \rstaticbalance as a network-flow problem.
As a major result in this regard, we show that solving \rstaticbalance on multigraph $\G$ is equivalent to solving the well-established \mincostflow problem~\cite{Ahuja93} on an ad-hoc modified version of $\G$.
We start by recalling the \mincostflow problem:
\begin{problem}[\mincostflow~\cite{Ahuja93}]\label{problem:mincostflow}
Given a simple directed graph $G=(V,E)$, a cost function $c: E \rightarrow \mathds{R}$, lower-bound and upper-bound functions $\lambda: E \rightarrow \mathds{R}$, $\mu: E \rightarrow \mathds{R}$, and a supply/demand function $b:V \rightarrow \mathds{R}$, find a flow $f: E \rightarrow \mathds{R}$ so as to
$$
\begin{array}{rl}
\mbox{Minimize} & \!\!\sum_{e \in E} c(e)f(e) \\
\mbox{subject to} & \!\!\lambda(e) \leq f(e) \leq \mu(e), \ \ \forall  e\in E \\
&  \!\!\!\!\!\!\!\!\!\!\!\!\! \!\!\!\!\!\!\sum _{u:(v,u)\in E}f(v,u) \!-\! \sum _{u:(u,v)\in E}f(u,v) = b(u), \forall u \in V
\end{array}
$$
\end{problem}
\noindent The modified version of $\G$ considered in this context is as follows:
\begin{definition}[\flowgraph]
The \emph{\flowgraph} $\G_f = (\V_f, \E_f, w_f)$ of an \bmultigraph $\G = (\V,\E,w)$ is a simple weighted directed graph where:
\begin{itemize}
\item
All arcs $(u,v) \in \E$ between the same pair of nodes are collapsed into a single one, and the weight $w_f(u,v)$ is set to $\sum_{(u,v) \in \E} w(u,v)$;
\item
$\V_f = \V \cup \{\tilde{s},\tilde{t}\}$, i.e., the node set of $\G_f$ is composed of all nodes of $\G$ along with two dummy nodes $\tilde{s}$ and $\tilde{t}$;
\item
$\E_f = \E \cup \{(\tilde{s}, u) \mid u \in \V\} \cup \{(u,\tilde{t}) \mid u \in V\} \cup \{(\tilde{t}, \tilde{s})\}$, i.e., the arc set of $\G_f$ is composed of ($i$) all (collapsed) arcs of $\G$, ($ii$) for each node $u \in \V$, a dummy arc $(\tilde{s},u)$ with weight $w_f(\tilde{s}, u) \!=\! \bala(u) \!-\! \lb(u)$ and a dummy arc $(u,\tilde{t})$ with weight $w_f(u,\tilde{t}) \!=\! \ub(u) \!-\! \balr(u)$, and ($iii$) a dummy arc $(\tilde{t}, \tilde{s})$ with weight $w_f(\tilde{t}, \tilde{s}) \!=\! \infty$.
\end{itemize}
\end{definition}

\noindent 
The main result for the computation of the desired upper bound is stated in the next theorem and corollary:
\begin{theorem}\label{theorem:flow}
Given an \bmultigraph $\G = (\V, \E, w)$, let $\G_f = (\V_f, \E_f, w_f)$ be the \flowgraph of $\G$.
Let also cost, lower-bound, upper-bound and supply/demand functions $c: \E_f \rightarrow \mathds{R}$, $\lambda: \E_f \rightarrow \mathds{R}$, $\mu: \E_f \rightarrow \mathds{R}$ and $b:\V_f \rightarrow \mathds{R}$ be defined as:
\begin{itemize}
\item
$\lambda(e) = 0$, $\mu(e) = w_f$, \ $\forall e \in \E_f$;
\item
$c(\tilde{t}, \tilde{s})=0$, and $c(\tilde{s},u) = c(u,\tilde{t}) = 0$, \ $\forall u \in \V_f$;
\item
$c(e) = -1$, \ $\forall e \in \E_f \cap \E$;
\item
$b(u) = 0$, \ $\forall u \in \V_f$.
\end{itemize}
It holds that solving {\normalfont\mincostflow} on input $\langle \G_f, c, \lambda, \mu, b \rangle$ is equivalent to solving {\normalfont\rstaticbalance} on input $\G$.
\end{theorem}
\begin{IEEEproof}
As $c(e) = -1$, if $e \in \E$, $c(e) = 0$, otherwise, and $\forall e \in \E:0 \leq f(e) \leq w(e)$,
the objective function of \mincostflow on $\G_f$ can be rewritten as $\min \sum_{e \in \E} -f(e)$, 
which is equivalent to the objective $\max \sum_{e \in \E} x_ew(e)$ ($x_e \in [0,1]$) of \rstaticbalance on $\G$.
For the cap-floor constraints, the conservation of flows ensures for any solution to \mincostflow on $\G_f$:
$$
\begin{array}{l}
\forall u \in \V: \\
 \sum_{(v,u) \in \E} \!f(v,u) \!+\! f(\tilde{s},u) \!-\! \sum_{(u,v) \!\in \E} \!f(u,v) \!-\! f(u, \tilde{t}) \!=\! b(u) \!=\! 0\\
\!\!\!\Leftrightarrow \!\forall u \!\in\! V\!:\!\! \sum_{(v,u) \in \E} f(v,u) - \sum_{(u,v) \in \E} f(u,v) =  f(u, \tilde{t}) - f(\tilde{s},u).
\end{array}
$$
As $f(\tilde{s},u) \in [f_{min}(\tilde{s},u), f_{max}(\tilde{s},u)] = [0, \bala(u) \!-\! \lb(u)]$ and $f(u, \tilde{t}) \in [f_{min}(u, \tilde{t}), f_{max}(u, \tilde{t})] = [0, \ub(u) \!-\! \balr(u)]$, then:
$$
\begin{array}{l}
\forall u \in \V: \sum_{(v,u) \in \E} f(v,u) - \sum_{(u,v) \in \E} f(u,v)\\
 = f(u, \tilde{t}) - f(\tilde{s},u)
\leq  f_{max}(u, \tilde{t}) - f_{min}(\tilde{s},u) \\
= \ub(u) - \balr(u),
\end{array}
$$
and
$$
\begin{array}{l}
\forall u \in \V : \sum_{(v,u) \in \E} f(v,u) - \sum_{(u,v) \in \E} f(u,v)\\
 = f(u, \tilde{t}) - f(\tilde{s},u) \geq  f_{min}(u, \tilde{t}) - f_{max}(\tilde{s},u)  \\
 =  \lb(u) - \bala(u).
\end{array}
$$
Overall, it is therefore guaranteed that $\forall u \in \V$:
$$
\sum_{(v,u) \in \E} \!\!\!f(v,u) -  \!\!\!\!\sum_{(u,v) \in \E}\!\!\f(u,v) \in [\lb(u) - \bala(u), \ub(u) - \balr(u)],
$$
i.e., the floor-cap constraints in \rstaticbalance.
\end{IEEEproof}

\begin{corollary}\label{corollary:flow}
Given an \bmultigraph $\G$, the solution to {\normalfont\staticbalance} on $\G$ is upper-bounded by the solution to {\normalfont\mincostflow} on the input $\langle \G_f, c, \lambda, \mu, b \rangle$ of \mbox{Theorem~\ref{theorem:flow}}.
\end{corollary}
\begin{IEEEproof}
Immediate as, according to Theorem~\ref{theorem:flow}, the solution to \textsc{Min-cost Flow} on input $\langle \G_f, c, \lambda, \mu, b \rangle$ corresponds to the optimal solution to a relaxed version of the \staticbalance problem on the original \bmultigraph $\G$.
\end{IEEEproof}

\begin{algorithm}[t]
\caption{\normalfont \bbalgub}
\label{alg:mincostflow}
\begin{algorithmic}[1]
\small
\REQUIRE An \bmultigraph $\G = (\V, \E, w)$, two multisets $\E_X^+ \subseteq \E$, $\E_X^- \subseteq \E$
\ENSURE A real number $UB_X$
\vspace{1mm}
\STATE $\!\G^- := (\V, \E \setminus \E_X^-, w)$
\STATE $UB_X \!\gets\! \mbox{solve \mincostflow  applying Theorem~\ref{theorem:flow}}$ on $\G^-\!$ \\
 and forcing flow $f(e) = w(e)$, $\forall e \in \E_X^+$; \\ 
 return $-\!1$ if no admissible solution exists
\end{algorithmic}
\end{algorithm}

\smallskip
Corollary~\ref{corollary:flow} is exploited for upper-bound computation as in Algorithm~\ref{alg:mincostflow}.
To handle the arcs to be discarded ($\E^-$) and included ($\E^+$), Algorithm~\ref{alg:mincostflow} removes all arcs $\E^-$ from the multigraph, and asks for a \mincostflow solution where the flow on every arc $e \in E^+$ is forced to be $w(e)$.
If no solution satisfying such a requirement exists, no admissible solution to \staticbalance exists in the entire subtree rooted in the target tree-node $X$. 
In this case the returned upper bound is $-1$, and the subtree rooted in $X$ is pruned by the \bbalg algorithm.
We solve \mincostflow with the well-established \textsf{Cost Scaling} algorithm~\cite{goldberg1990finding}, which has $\bigO(|\E|~\!(|\V|\log|\V|)~\!\log(|\V|~\!w_{max}))$ time complexity, where $w_{max} = \max_{e \in \E} w(e)$.
This also corresponds to the time complexity of the entire upper-bound computation method.

\subsection{Beam-search algorithm}\label{section:approxalg}

Being \staticbalance \NPhard, the exact \bbalg algorithm cannot handle large \bmultigraph{s}.
We thus design an alternative algorithm that finds approximate solutions and can run on larger instances.
To this end, we exploit the idea of enumerating all cycles in the input multigraph, and selecting a subset of them while keeping the constraints satisfied. 
The intuition behind this strategy is twofold.
First, a solution composed of a set of cycles always satisfies the other constraint of the problem, as every node of a cycle has at least one incoming arc and one outgoing arc.
Also, cycle enumeration is a well-established problem, for which a variety of algorithms exist.
As a trade-off between effectiveness, efficiency and simplicity, in this work we employ a variant of the Johnson's algorithm~\cite{Johnson75} that works on multigraphs~\cite{Hawick08}.

\spara{Optimal cycle selection.} 
For a principled cycle selection, we focus on the problem of seeking cycles that satisfy the constraints in Problem~\ref{problem:balancestatic} and exhibit maximum~total~amount:
\begin{problem}[\optcyclesel]\label{prob:optcyclesel}
\!\!\!Given an \bmultigraph $\G \!=\! (\V,\E,w)$ and a set $\cycles$ of cycles in $\G$, find a subset $\cycles^* \!\subseteq\! \cycles$ so~that:
$$
\vspace{-1mm}
\begin{array}{rl}
\!\!\!\!\!\!\!\!\!\!\!\!\cycles^* = & \!\!\!\arg\max_{\hat{\cycles} \subseteq \cycles}  \sum_{e \in \E(\hat{\cycles})} w(e) \\
\!\!\!\!\!\!\!\!\!\!\!\!\mbox{subject to} & \!\!\!\mbox{$\E(\hat{\cycles}) \!=\! \bigcup_{C \in \hat{\cycles}} C$ meets Constraint~(\ref{eq:prob1constraint1}) in Problem~\ref{problem:balancestatic}}.
\end{array}
$$
\end{problem}

\begin{theorem}\label{th:optcycleselhardness}
Problem~\ref{prob:optcyclesel} is \NPhard.
\end{theorem}
\begin{IEEEproof}
We reduce from \maxindset~\cite{garey1979computers}, which asks for a maximum-sized subset of vertices in a graph no two of which are adjacent.
Given a graph $G = (V,E)$ instance of \maxindset, we construct an instance $\langle \G, \cycles\rangle$ of \optcyclesel as follows.
For a vertex $u \in V$, let $N(u) = \{v \mid (u,v) \in E\}$,
$d_u = |N(u)|$, $d_{max} = \max_{u \in V} d_u$.
Without loss of generality, we assume that $\forall u \in V:d_u > 0$.
First, we let the node set $\V$ of $\G$ contain: ($i$) a pair of nodes $\langle u', u'' \rangle$ for every vertex $u \in V$, ($ii$) a node $uv$ for each $(u,v) \in E$. 
Let also $\V$ follow a global ordering.
For each $u \in V$, we create a cycle $C_u = c_1 \rightarrow c_2 \rightarrow \cdots \rightarrow c_{d_u+2} \rightarrow c_1$ in $\G$, where $c_1 = u'$, $c_2=u''$, and all other nodes $c_i:i>2$  correspond to nodes $\{uv \in \V \mid v \in N(u)\}$, 
ordered as picked above. 
The arc weights of $C_u$ are defined as follows:
\begin{itemize}
\item
$w(c_1,c_2) = 1 + \frac{3}{2}[d_{max}(d_{max}+1) - d_u(d_u+1)]$,
\item
$w(c_i, c_{i+1}) = d_u + i - 2$, $\forall i \in [2..d_u+1]$,
\item
$w(c_{d_u+2},c_1) = 2d_u$.
\end{itemize}
Finally, we set 
\begin{itemize}
\item
$\balr(x) = \bala(x) = 0$, \ $\ub(x) = +\infty$, \ $\forall x \in \V$,
\item
$\lb(u') = \lb(u'') = -\infty$, $\forall u',u''$,
\item
$\lb(uv) = -1$, $\forall uv$, 
\end{itemize}
and the input cycles $\cycles$ equal to $\{C_u \mid u \in V\}$. It holds that:
\begin{enumerate}
\item[(a)]
$\G$ has $2|V| \!+\! |E|$ nodes and $\sum_{u \in V'} (d_u\!+\!2) = 2(|V| \!+\! |E|)$ arcs, taking polynomial space and construction time in the $G$'s size.
\item[(b)]
All cycles in $\cycles$ are arc-disjoint with respect to each other.
\item[(c)]
Every cycle $C_u \in \cycles$ has the same total amount, which is  equal to 
$1 + \frac{3}{2}[d_{max}(d_{max}+1) - d_u(d_u+1)]$ 
$ + \sum_{i=0}^{d_u} (d_u+i)$ 
$ =  1 + \frac{3}{2}[d_{max}(d_{max}\!+\!1) - d_u(d_u\!+\!1)] + \frac{3}{2}d_u(d_u+1)$
$ \!=\! 1 + \frac{3}{2}d_{max}(d_{max}\!+\!1)$.
\item[(d)]
Selecting any two cycles $C_u, C_v \in \cycles$ 
such that $u$ and $v$ are adjacent in $G$,
violates the constraint on $\lb(uv)$ (as $\bala(uv)$ will result to be $\bala(uv) = -2 < \lb(uv) = -1)$).
\end{enumerate}
Based on (b) and (c), any cycle brings the same gain to the objective.
Thus, solving \optcyclesel on $\langle \G, \cycles \rangle$ corresponds to selecting the \emph{maximum number} of cycles in $\cycles$ so that cap-floor constraints are met.
Combined with (d), solving \optcyclesel on $\langle \G, \cycles \rangle$ is equivalent to selecting the maximum number of vertices in $G$ no two of which are adjacent. 
\end{IEEEproof}

\begin{algorithm}[t]
\caption{\normalfont \bsalg}
\label{alg:bsalg}
\begin{algorithmic}[1]
\small
\REQUIRE \bmultigraph $\G = (\V, \E, w)$, \ integer $K$
\ENSURE Multiset $\E^* \subseteq \E$
\vspace{0mm}
\STATE $\E^* \gets \emptyset$
\STATE $\cycles \gets \mbox{cycles of } \G$ \hfill\COMMENT{\cite{Hawick08}}
\WHILE{$\cycles \neq \emptyset$}
\STATE $\cycles' \gets \mbox{$K$-sized subset of $\cycles$ by \textsf{Greedy} \maxcover}$\hfill\COMMENT{\cite{hochbaum1996approximating}}
\STATE $\cycles'_2 \gets \{\{C_i, C_j\} \!\mid\! C_i, C_j \in \cycles'\}$
\FORALL{$\{C_i, C_j\} \in \cycles'_2$}
\STATE $C_{ij} \gets C_i \cup C_j$
\STATE process all $C \!\in\! \cycles' \!\setminus\! \{C_i, C_j\}$ one by one, by non-increasing $\cyclescoreSUK(\cdot)$ score (Eq.(\ref{eq:cyclescoreSUK}));
add $C$ to $C_{ij}$ if $C_{ij} \cup C \cup \E^*$ is feasible for Problem~\ref{prob:optcyclesel}
\ENDFOR
\STATE $\E^* \gets \E^* \cup \arg\max_{C_{ij} \in \cycles'_2} \sum_{e \in C_{ij}} w(e)$
\STATE $\cycles \gets \cycles \setminus (\cycles' \cup \{C \in \cycles \mid C \cap \E^* = C\})$
\ENDWHILE
\end{algorithmic}
\end{algorithm}

\spara{Characterization as a \knapsack problem.}
As \optcyclesel is \NPhard, we focus on designing effective approximated solutions.
To this end, we show an intriguing connection with the following variant of the well-known \knapsack problem:

\begin{problem}[\suk~\cite{goldschmidt1994note}]\label{problem:suk}
Let  $U = \{x_1, \ldots, x_h\}$ be a universe of elements,
$\mathcal{S} = \{S_1, \ldots, S_k\}$ be a set of items, where $S_i \subseteq U$, $\forall i \in [1..k]$,
 $p: \mathcal{S} \rightarrow \mathds{R}$ be a profit function for items in $\mathcal{S}$,
and  $q: U \rightarrow \mathds{R}$ be a cost function for elements in $U$.
For any $\hat{\mathcal{S}} \subseteq \mathcal{S}$ define also: 
$U(\hat{\mathcal{S}}) = \bigcup_{S \in \hat{\mathcal{S}}}S$,
$P(\hat{\mathcal{S}}) = \sum_{S \in \hat{\mathcal{S}}} p(S)$,
and $Q(\hat{\mathcal{S}}) = \sum_{x \in U(\hat{\mathcal{S}})} q(x)$.
Given a real number $B \in \mathds{R}$, \suk finds \
$
\mathcal{S}^* = \arg\max_{\hat{\mathcal{S}} \subseteq \mathcal{S}} P(\hat{\mathcal{S}}) \ \  \mbox{s.t.} \ \ Q(\hat{\mathcal{S}}) \leq B. 
$
\end{problem}

A simple variant of \suk arises when both costs and budget constraint are $d$-dimensional:

\begin{problem}[\multisuk]\label{problem:multisuk}
Given $U$, $\mathcal{S}$, $p$ as in Problem~\ref{problem:suk}, 
a $d$-dimensional cost function $q : U \rightarrow \mathds{R}^d$, and a $d$-dimensional vector $\vec{B} \in \mathds{R}^d$,
find 
$
\mathcal{S}^* = \arg\max_{\hat{\mathcal{S}} \subseteq \mathcal{S}} P(\hat{\mathcal{S}}) $  
$
\mbox{s.t.} \ \ \vec{Q}(\hat{\mathcal{S}}) \leq \vec{B}, 
$
where $\vec{Q}(\hat{\mathcal{S}}) = \sum_{x \in U(\hat{\mathcal{S}})}q(x)$.
\end{problem}
We observe that an instance of \optcyclesel can be transformed into an instance of \multisuk so that every feasible solution for the latter instance is also a feasible solution for the original \optcyclesel instance.
We let the arcs $\E$ represent elements, cycles $\cycles$ represent items, and define $2|\V|$ costs/budgets  for each element (arc), so as to match the cap-floor constraints on every node ($|\V|$ costs/budgets for cap- and floor-related constraints each).
Formally:
 %
\begin{theorem}\label{theorem:multisukVSoptcyclesel}
Given an \bmultigraph $\G = (\V, \E, w)$ and a set $\cycles$ of cycles in $\G$, let $\langle U, \mathcal{S}, p, q, \vec{B} \rangle$ be an instance of {\normalfont\multisuk} defined as follows:
\begin{itemize}
\item
$U = \E$; \ $\mathcal{S} = \cycles$; \ $\forall C \in \cycles: p(C) = \sum_{e \in C} w(e)$;
\item
$\forall (u_i,u_j) \!\in\! \E: q(u_i,u_j) \!=\! [\vec{q}^+(u_i,u_j)~\!\vec{q}^-(u_i,u_j)] \!\in\! \mathds{R}^{2|\V|}$, where
\begin{itemize}
\item
$\vec{q}^+(u_i,u_j) = [q^+_k(u_i,u_j)]_{k=1}^{|\V|}$, with $q^+_i(u_i,u_j) = -w(u_i,u_j)$, $q^+_j(u_i,u_j) = w(u_i,u_j)$, and $q^+_k(u_i,u_j) = 0$, for $k \neq i,j$;
\item
$\vec{q}^-(u_i,u_j) = [q^-_k(u_i,u_j)]_{k=1}^{|\V|}$, with $q^-_i(u_i,u_j) = w(u_i,u_j)$, \\ $q^-_j(u_i,u_j)~=~-w(u_i,u_j)$, and $q^-_k(u_i,u_j) = 0$, for $k \neq i,j$;
\end{itemize}
\item
$\vec{B} := [\vec{B}^+ \ \vec{B}^-] \in \mathds{R}^{2|\V|}$, where 
\begin{itemize}
\item
$\vec{B}^+ = [\ub(u_1) - \balr(u_1), \ldots, \ub(u_n) - \balr(u_n)]$;
\item
$\vec{B}^- = [\bala(u_1) - \lb(u_1), \ldots, \bala(u_n) - \lb(u_n)]$.
\end{itemize}
\end{itemize}
\vspace{1mm}
\noindent
It holds that any feasible solution for {\normalfont\multisuk} on input $\langle U, \mathcal{S}, p, q, \vec{B} \rangle$ is a feasible solution for {\normalfont\optcyclesel} on input $\langle \G, \cycles\rangle$.
\end{theorem}
\begin{IEEEproof}
(\textsc{Sketch}) It suffices to show that the constraints of \multisuk on $\langle U, \mathcal{S}, p, q, \vec{B} \rangle$ correspond to cap-floor constraints of \optcyclesel on $\langle \G, \cycles\rangle$.
This can be achieved by simple math on $\vec{q}^+$\!\!, $\vec{q}^-$\!\!, $\vec{B}^+$\!\!, $\vec{B}^-$\!\!.
\end{IEEEproof}

\spara{Putting it all together.}
Motivated by Theorem~\ref{theorem:multisukVSoptcyclesel}, we devise an algorithm to approximate \optcyclesel inspired by Arulselvan's algorithm for \suk~\cite{arulselvan2014note}, which achieves a $1 \!-\! e^{-1/f_{max}}$ approximation guarantee, where $f_{max}$ is the maximum number of items in which an element is present.
Arulselvan's algorithm considers all subsets of 2 items whose weighted union is within the budget $B$.
Then, it augments each subset with items $S_i$ added one by one in the decreasing order of an ad-hoc-defined score, as long as the inclusion of $S_i$ complies with the budget constraint $B$.
The score exploited for item processing is directly proportional to the profit of $S_i$ and inversely proportional to the frequency of $S_i$'s elements within the entire item set $\mathcal{S}$. 
The highest-profit one of such augmented subsets is returned as output.

The ultimate \bsalg algorithm (Algorithm~\ref{alg:bsalg}) combines the ideas behind Arulselvan's algorithm with the \emph{beam-search} paradigm and a couple of adaptations to make it suitable for our context.
Particularly, the adaptations are as follows.
($i$) We extend Arulselvan's algorithm so as to handle a \multisuk problem instance derived from the input \optcyclesel instance as stated in Theorem~\ref{theorem:multisukVSoptcyclesel} (trivial extension).
($ii$) We define the score of a cycle $C \in \cycles$ as:
\vspace{-1mm}
\begin{equation}\label{eq:cyclescoreSUK}
\cyclescoreSUK(C) = \frac{\sum_{e \in C} w(e)}{\sum_{e \in C} \frac{w(e)}{f(e)}}, \ \mbox{where } f(e) = |\{C \in \cycles : e \in C\}|,\vspace{-1mm}
\end{equation}
whose rationale is to consider the total cycle amount, penalized by a term accounting for the frequency of an arc within the whole cycle set $\cycles$.
The idea is that frequent arcs contribute less to the objective function, which is defined on the \emph{union} of the arcs of all selected cycles (see Problem~\ref{prob:optcyclesel}).
Finally, ($iii$) to overcome the expensive pairwise cycle enumeration and augmentation of \emph{all} 2-sized cycle sets with \emph{all} other cycles, we adopt a beam-search methodology to reduce the cycles to be considered.
We first select a subset $\cycles' \subseteq \cycles$ of cycles of size $K \leq |\cycles|$ whose union arc set exhibits the maximum amount,\footnote{To solve this step, we note that the problem at hand is an instance of (weighted) \maxcover~\cite{hochbaum1996approximating}, where arcs correspond to elements and cycles correspond to sets. We hence employ the classic greedy $(1 - \frac{1}{e})$-approximation algorithm, which iteratively adds to the solution the cycle maximizing the sum of the weights of still uncovered arcs.}
and use $\cycles'$ for both pairwise cycle computation and augmentation.
We repeat the procedure (by selecting a further $K$-sized subset of $\cycles \setminus \cycles'$) until $\cycles$ has become empty.

\spara{Time complexity.}
As far as cycle enumeration, the considerations made for Algorithm~\ref{alg:johnson} (Section~\ref{sec:bbalg}) remain valid here too.
The complexity of the remaining main steps is as follows.
Denoting by $L$ the maximum size of a cycle in $\cycles$, the greedy \maxcover algorithm on input $\cycles'$ (Line~4) takes $\bigO(L~\!K\log K)$ time.
All feasible cycle pairs (Line~5) can be computed in $\bigO(L~\!K^2)$ time, while augmentation of all such pairs (Lines~6--8) takes $\bigO(L~\!K^3)$ time.
All these steps are repeated $\bigO(\frac{|\cycles|}{K})$ times. The overall time complexity is $\bigO(L~\!K^2~\!|\cycles|)$.

\begin{algorithm}[t]
\caption{\normalfont \pathalg}
\label{alg:pathalg}
\begin{algorithmic}[1]
\small
\vspace{0mm}
\REQUIRE \bmultigraph $\G = (\V, \E, w)$, \ two integers $K, K_p$
\ENSURE Multiset $\E^* \subseteq \E$
\vspace{0mm}
\STATE $\cycles^* \gets $ cycles of $\G$ selected by Algorithm~\ref{alg:bsalg}
\STATE $\E^* \gets \bigcup_{C \in \cycles^*} C$
\FORALL{$C_i, C_j \in \cycles^*$}
	\STATE $\paths_{ij} \gets $ paths from a node in $C_i$ to a node in $C_j$\hfill\COMMENT{\cite{Hawick08}}
	\STATE $\paths_{ij}^* \gets $ paths selected by Algorithm~\ref{alg:bsalg} on input $\langle \G, K_p \rangle$ and setting $\cycles := \paths_{ij}$ at Line~2
	\STATE $\E^* \gets \E^* \cup \bigcup_{P \in \paths^*_{ij}} P$
\ENDFOR
\end{algorithmic}
\end{algorithm}

\subsection{Adding paths between cycles}
\label{algo:paths}

An interesting insight on cycle-based solutions to \staticbalance is that they can be improved by \emph{looking for paths connecting two selected cycles}.
An approach based on this observation is particularly appealing as adding a path between two cycles keeps the cycle constraint of our problem (i.e., Constraint~(2) in Problem~\ref{problem:balancestatic}) satisfied, therefore one has only to focus on the floor-cap constraint while checking admissibility of paths.

Within this view, we propose a refinement of the \bsalg algorithm where, for every pair $C_i, C_j$ of selected cycles, we ($i$) compute paths $\paths_{ij}$ from a node in $C_i$ to a node in $C_j$, and ($ii$) take a subset  $\paths^*_{ij} \subseteq \paths_{ij}$ such that the total amount $\sum_{e \in P, P \in \paths^*_{ij}} w(e)$ is as large as possible and  adding $\paths^*_{ij}$ to the initial \staticbalance solution still meets the cap-floor constraints.
We dub this algorithm \pathalg and outline it as Algorithm~\ref{alg:pathalg}.
To compute a path set $\paths_{ij}$, we employ a simple variant of the (multigraph version of the) well-known Johnson's algorithm~\cite{Hawick08}, where we make the underlying backtracking procedure start from nodes in $C_i$  and yield a solution (i.e., a path) every time it encounters a node in $C_j$.
The selection of $\paths^*_{ij} \subseteq \paths_{ij}$ can be accomplished by running either a standard greedy method (as in Algorithm 2 in the first version of our work~\cite{bordino18network}), or Algorithm~\ref{alg:bsalg} (with the modification that the set $\cycles$ at Line~2 corresponds to $\paths_{ij}$).

The running time of \pathalg is expected to be dominated by the various path-enumeration steps.
As far as the remaining steps, the worst case corresponds to employing Algorithm~\ref{alg:bsalg} for both cycle- and path-selection, which leads to an overall time complexity of $\bigO(L K^2 |\cycles| + |\cycles^*|^2 L_p K_p^2 |\paths_{max}|)$, where $L_p$ and $K_p$ are two integers playing the same roles as $L$ and $K$, respectively, in Algorithm~\ref{alg:bsalg} when run on path sets, and $\paths_{max} = \argmax_{\paths \in \{\paths_{ij}\}_{i,j}} |\paths|$.

\subsection{Hybrid algorithm}
\label{algo:hybrid}
The \staticbalance problem can be solved by running any of the algorithms presented in Sections~\ref{sec:bbalg}--\ref{algo:paths} on each weakly connected component of the input \bmultigraph \emph{separately}, and then taking the union of all partial solutions.
Based on this straightforward observation, we ultimately propose a \emph{hybrid} algorithm, which runs the exact \bbalg algorithm on the smaller connected components and the \bsalg or \pathalg algorithm on the remaining ones.

Our hybrid algorithm is outlined as Algorithm~\ref{alg:hybridalg}: we term it either \mbox{\hybridalg} or \hybridalgp, depending on whether it is equipped with \bsalg or \pathalg, respectively.

\begin{algorithm}[t]
\caption{\normalfont \hybridalg/\hybridalgp}
\label{alg:hybridalg}
\begin{algorithmic}[1]
\small
\vspace{0mm}
\REQUIRE \bmultigraph $\G = (\V, \E, w)$, \ three integers $H, K, K_p$
\ENSURE Multiset $\E^* \subseteq \E$
\vspace{0mm}
\STATE $\E^* \gets \emptyset$, \ \ $\mathbf{CC} \gets \mbox{weakly connected components of }\G$
\FORALL{$G \in \mathbf{CC}$ s.t. $|arcs(G)| \leq H$}
\STATE $\E^* \gets \E^* \cup $ \bbalg on input $G$\hfill\COMMENT{Section~\ref{sec:bbalg}}
\ENDFOR
\FORALL{$G \in \mathbf{CC}$ s.t. $|arcs(G)| > H$}
\STATE $\E^* \gets \E^* \cup $ \bsalg or \pathalg on input $\langle G, K, K_p \rangle$\hfill\COMMENT{Algorithms~\ref{alg:bsalg}--\ref{alg:pathalg}}
\ENDFOR
\end{algorithmic}
\end{algorithm}

\subsection{Temporary constraint violation}
\label{algo:negbal}

Given a set $\E^*$ of receivables selected for settlement, the eventual step of our network-based RF service is, for every receivable $R \in \E^*$, to transfer $\amount(R)$ from $\debtor(R)$'s account to $\creditor(R)$'s account.
A typical desideratum from big (financial) institutions is that these money transfers happen by exploiting procedures and software solutions already in place within the institution, in order to minimize the effort in system re-engineering.
For instance, our partner institution required to interpret money transfers as a sequence of standard bank transfers. 
To accomplish this, money transfers have to typically be executed one at a time,
meaning that \emph{some constraints in Problem~\ref{problem:balancestatic} might be temporarily violated}.
As an example, look at Figure~\ref{img:example} and 
assume the money transfer corresponding to the receivable from \textsf{D} to \textsf{E} is the first one to be executed: this would lead to the temporary violation of \textsf{D}'s floor constraint. 
In real scenarios constraint violation might not be allowed, not even temporarily.
In banks, for instance, a floor-constraint violation corresponds to an overdraft on a bank account, which, even if it lasts a few milliseconds only, would anyway cause that account being charged.
Another motivation may be simply that existing systems do not accept any form of temporary violation.

Motivated by the above, here we devise solutions to the temporary-constraint-violation issue.
In particular, we focus on floor constraints only, as temporary violation of cap constraints or cycle constraints is not really critical.
We propose two solutions: the first one is based on a clever redefinition  of floor values, while the second one consists in properly selecting and ordering a subset of the arcs that were originally identified for settlement.
The details of both solutions are reported next.

\spara{Redefining floor values.}
The first proposal to handle temporary constraint violation is based on the following result: properly redefining floor values leads to output solutions to \staticbalance containing only receivables that do not violate the original floor constraints, \emph{regardless of the ordering with which the corresponding money transfers are executed}.
Specifically, given a set $\E^* \subseteq \E$ of arcs in a \staticbalance solution, for every node $u \in \V(\E^*)$, let $in(u,\E^*) = \sum_{v:(v,u) \in \E^*}w(v,u)$ and $out(u,\E^*) = \sum_{v:(u,v) \in \E^*} w(u,v)$ denote the total amount received and paid by $u$ within $\E^*$, respectively.
Let also $ub_{in}(u)$ denote an upper bound on the amount $in(u,\E^*)$ that $u$ can receive in any possible solution $\E^*$. 
Our first observation is that, by setting $\lb(u) = ub_{in}(u)$, $\forall u \in \V(\E^*)$, no customer $u$ will ever pay a total amount larger than her current $\bala(u)$ balance availability $\bala(u)$ in any \staticbalance solution, i.e., it would be guaranteed that, for all solutions $\E^*$ and all $u  \in \V(\E^*)$, $\bala(u) \geq out(u,\E^*)$.
In fact, for all solutions $\E^*$ it holds that:
\begin{eqnarray}
\lefteqn{in(u,\E^*) - out(u,\E^*) + \bala(u) \ \geq \  \lb(u) = ub_{in}(u)}\nonumber\\
& \Leftrightarrow & \bala(u) - out(u,\E^*) \ \geq \ ub_{in}(u) - in(u,\E^*) \ \geq \ 0\nonumber\\
& \Rightarrow & \bala(u) \ \geq \ out(u,\E^*)\label{eqtmp1}.
\end{eqnarray}
Now, the key question is how to define $ub_{in}(u)$.
A simple way would be setting it  to the $u$'s in-degree.
Another option consists in defining it based on the cap constraint:
\begin{eqnarray}
\lefteqn{\ub(u) \ \geq \ in(u,\E^*) \!-\! out(u,\E^*) \!+\! \balr(u)}\nonumber\\
& \Rightarrow & \ub(u) \ \geq \ in(u,\E^*) \!-\! \bala(u) \!+\! \balr(u) \quad \mbox{\{Equation~(\ref{eqtmp1})\}}\nonumber\\ 
& \Leftrightarrow & in(u) \ \leq \ \ub(u) \!+\! \bala(u) \!-\! \balr(u).\nonumber
\end{eqnarray}
Combining the two options, we ultimately define
\begin{equation}\label{eqtmp3}
\!ub_{in}\!(u) \!=\! \min \!\Big \{\! {\textstyle\sum_{v:(v,u) \in \E} \!w(v,\!u),  \ub(u) \!+\! \bala\!(u) \!-\! \balr\!(u)} \!\Big\}\!.\!\!\!\!\!
\end{equation}

\begin{algorithm}[t]
\caption{\normalfont \negbalalg}
\label{alg:negbal}
\begin{algorithmic}[1]
\small
\vspace{0mm}
\REQUIRE \bmultigraph $\G = (\V, \E, w)$, \ two integers $H, K$
\ENSURE List $\E^* \subseteq \E$
\vspace{0mm}
\STATE $\E^* \gets [~\!]$
\REPEAT
	\STATE $\widehat{\E} \gets $ a solution to Problem~\ref{problem:balancestatic} on input $\langle \G, H, K \rangle$, with floors set as $\lb(u) = ub_{in}(u)$, $\forall u \in \V$\hfill\COMMENT{Equation~(\ref{eqtmp3})}
	\STATE order  $\widehat{\E}$ in some way and append it to $\E^*$
	\STATE remove all arcs in $\widehat{\E}$ from $\G$
	\FORALL{$u \in \V(\widehat{\E})$}
		\STATE $\Delta(u) \gets \sum_{v:(v,u) \in \widehat{\E}} w(v,u) - \sum_{v:(u,v) \in \widehat{\E}} w(u,v)$
		\STATE $\balr(u) \gets \balr(u) + \Delta(u)$, \ $\bala(u) \gets \bala(u) + \Delta(u)$
	\ENDFOR
\UNTIL{$\E = \emptyset \ \vee \ \widehat{\E} = \emptyset$}
\end{algorithmic}
\end{algorithm}

To summarize, our first strategy to handle temporary constraint violation consists in computing a solution $\E^*$ to \staticbalance with a floor constraint $\lb(u) = ub_{in}(u)$ (Equation~(\ref{eqtmp3})), for all $u$.
Based on the above arguments,  $\E^*$ guarantees no temporary floor-constraint violation, independently of the execution order of the corresponding money transfers.
As a refinement, one can run this method iteratively, i.e.,  removing the arcs yielded at any iteration and computing a new solution on the remaining graph.
The overall solution is the union of all partial solutions, and the arcs are ordered based on the iterations:
money transfers corresponding to arcs computed in earlier iterations should be executed before the ones in later iterations (while arc ordering within the same iteration does not matter).
This method is outlined as Algorithm~\ref{alg:negbal}.

\begin{algorithm}[t]
\caption{\normalfont \negbalalgtwo}
\label{alg:negbal2}
\begin{algorithmic}[1]
\small
\vspace{0mm}
\REQUIRE \bmultigraph $\G = (\V, \E, w)$, \ two integers $H, K$
\ENSURE List $\E^* \subseteq \E$
\vspace{0mm}
\STATE $\E^* \gets [~\!]$; \ \ $i \gets 1$
\REPEAT
	\STATE $\E^+ \gets \emptyset$
	\FORALL{$u \in \V(\E)$}
		\STATE $\E_u \gets $ a subset of $\{(u,v) \mid (u,v) \in \E\}$ s.t. $\bala(u) - \sum_{e \in \E_u} w(e) \geq \lb(u)$, $\forall (u,v) \in \E_u : w(u,v) + \balr(v) \leq \ub(v)$, and $\sum_{e \in \E_u} w(e)$ is maximized
		\STATE $\E^+ \gets \E^+ \cup \E_u$
	\ENDFOR
	\STATE set $ts(e) = i$, $\forall e \in \E^+$
	\STATE order  $\E^+$ in some way and append it to $\E^*$
	\FORALL{$u \in \V(\E^+)$}
		\STATE $\Delta(u) \gets \sum_{v:(v,u) \in \E^+} w(v,u) - \sum_{v:(u,v) \in \E^+} w(u,v)$
		\STATE $\balr(u) \gets \balr(u) + \Delta(u)$, \ $\bala(u) \gets \bala(u) + \Delta(u)$
	\ENDFOR
	\STATE $\E \gets \E \setminus \E^+$; \ \ $i \gets i +1$
\UNTIL{$\E^+ = \emptyset$}
\STATE $\E^- \gets \emptyset$
\REPEAT
	\STATE $\E^- \gets \E^* \setminus \{$arcs of the (1,1)-D-core of $\E^*\}$\hfill\COMMENT{\cite{Giatsidis2011}}
	\FORALL{$(u,v) \in \E^-$}
		\STATE $\E^-_{uv} \gets \{(u,v)\}$; \ \ $X \gets \{(u,v)\}$
		\WHILE{$X \neq \emptyset$}
			\STATE $\widehat{X} \gets \emptyset$
			\FORALL{$(x,y) \in X$}
				\STATE $\widehat{X} \gets \widehat{X} \cup \{(y,z) \in \E^* \mid ts(y,z) = ts(x,y) + 1\}$
			\ENDFOR
			\STATE $X \gets \widehat{X}$; \ \ $\E^-_{uv} \gets \E^-_{uv} \cup X$
		\ENDWHILE
		\STATE remove all arcs in $\E^-_{uv}$ from $\E^*$
	\ENDFOR
\UNTIL{$\E^- = \emptyset$}
\end{algorithmic}
\end{algorithm}

\begin{table*}[t]
\vspace{-0mm}
\begin{center}
\vspace{-0mm}
{
\caption{{\footnotesize \label{tbl:graph-stats} Size of the input \bmultigraph{s} extracted from the real dataset considered in the experiments (algorithm: \hybridalg).}} 
\vspace{-0.2cm}
\scriptsize
\begin{tabular}{@{}c@{ }@{ }c@{ }||c@{ }||S[table-format=5]@{~~}S[table-format=6]|S[table-format=5]@{~~}S[table-format=6]||S[table-format=5]@{~~}S[table-format=6]|S[table-format=5]@{~~}S[table-format=6]||S[table-format=6]@{~~}S[table-format=6]|S[table-format=6]@{~~}S[table-format=6]@{}} 
\multirow{3}{*}{\textbf{dataset}} & \multicolumn{1}{c||}{\multirow{3}{*}{\textbf{span}}} & \multicolumn{1}{c||}{\multirow{3}{*}{\textbf{size}}}  & \multicolumn{4}{c||}{\textbf{Worst scenario}} & \multicolumn{4}{c||}{\textbf{Normal scenario}} & \multicolumn{4}{c}{\textbf{Best scenario}} \\
 & & & \multicolumn{2}{c|}{\textbf{cap} $< \infty$} & \multicolumn{2}{c||}{\textbf{cap} $= \infty$} & \multicolumn{2}{c|}{\textbf{cap} $< \infty$} & \multicolumn{2}{c||}{\textbf{cap} $= \infty$} & \multicolumn{2}{c|}{\textbf{cap} $< \infty$} & \multicolumn{2}{c}{\textbf{cap} $= \infty$}\\
 & & & \textbf{avg} & \textbf{max} & \textbf{avg} & \textbf{max}  & \textbf{avg} & \textbf{max} & \textbf{avg} & \textbf{max} & \textbf{avg} & \textbf{max} & \textbf{avg} & \textbf{max}\\
\hline
\multirow{2}{*}{\firstlog} &  {\multirow{2}{*}{\textbf{2015/16}}}  & \textbf{\#nodes} &  40810  &  58390 &  40810 &  58380  &  62890 & 82930  &  62870 &  82900  &   80260  &  97400   &  80230 &   97380 \\ 
& & \textbf{\#arcs} &  41720 &   69270  &  41670  &   69180   & 76490  &  116000  &  76310  & 115600   &  111100 & 149300  &  110700 &  148500 \\ 
\hline 
\multirow{2}{*}{\secondlog} &  {\multirow{2}{*}{\textbf{2017/18}}}  & \textbf{\#nodes} &   50563  & 77323   &  50548   &  77280  & 77134  & 105671   &  77082  & 105627 & 97888 & 121274 &  97789   & 121244   \\
& & \textbf{\#arcs}  & 51708   &  132944  &  51600 & 132786  &  92560 & 215913 &  92025  & 215078 & 132564  & 278246 & 131325  & 276497 \\
\end{tabular}}
\end{center}
\vspace{-0mm}
\end{table*}

\spara{Taking a subset of arcs and properly ordering them.}
Our second solution to overcome  temporary constraint violation consists in \emph{selecting a subset} of the arcs that were originally identified for settlement,  and \emph{ordering} them, so as to guarantee that ($i$) executing the corresponding money transfers according to that ordering is free from constraint violations, and ($ii$) the arc subset still satisfies \staticbalance's constraints.
While doing so, the  objective remains \emph{maximizing the total amount} of the arcs in the subset.

More specifically, the algorithm at hand (Algorithm~\ref{alg:negbal2}) consists of two phases, i.e., arc selection (Lines~2--13) and arc removal (Lines~14--25).
Arc selection
 aims at identifying, for every node $u$, the best (in terms of overall amount) subset $\E_u$ of arcs  outgoing from $u$ whose total amount complies with $u$'s floor constraint, and such that no arc $(u,v) \in \E_u$ leads to the violation of  $v$'s cap constraint  (Line~5).
 This corresponds to (a simple variant of) the \subsetsum problem, which is known to be \NPhard~\cite{Kellerer04}. 
For nodes with small-sized out-neighborhood, the problem might even be solved optimally, by brute-force.
Alternatively, for nodes with larger out-neighborhoods, an approximated solution can be computed by adopting some existing approximation algorithms for \subsetsum.
Once such $\E_u$ subsets, for all $u \in \V(\E)$, have been identified, all arcs within such $\E_u$ sets are appended to the output $\E^*$ list, with their timestamp set as equal to the current $i$ (Lines~7--8), and the balances of the corresponding nodes are updated (Lines~9--11).
As the selection of some arcs during a certain iteration may increase the balance of some nodes, and, thus, enable the selection of further arcs in the next iterations, the arc-selection phase is repeated until no new arc is selected in the current iteration.

The (temporary) list of arcs that has been built during arc selection guarantees that the floor-cap constraints of every selected node are still satisfied.
Moreover, if the corresponding money transfers are executed following the ordering of the list, it is also ensured that no floor constraint will ever be violated.
On the other hand, the current solution may violate the second constraint of the \staticbalance problem, i.e., the one that every node within the solution has at least one incoming arc and at least one outgoing arc.
Hence, the algorithm has also an arc-removal phase, where those arcs $\E^-$ that are not part of the (1,1)-D-core~\cite{Giatsidis2011} of the subgraph induced by the current solution are discarded (Line~16), and arcs that have been selected ``thanks to'' such discarded arcs are (iteratively) removed too (Lines~17--24). 
Arc removal is iteratively repeated until an empty $\E^-$ set has been built step, as removing arcs in one iteration may cause further constraint violations.

\section{Implementation}
\label{sec:implementation}

Here we provide some insights for a correct and/or efficient implementation of the proposed algorithms.

As a first observation, all algorithms can benefit from a preprocessing step, where nodes with no incoming or outgoing arcs are filtered out of the input \bmultigraph.
In fact, the removal of those nodes is safe as they certainly violate Constraint~(\ref{eq:prob1constraint2}) in our \staticbalance problem.
Such a filtering may be exploited recursively, as the removal of those nodes may cause further nodes to have no incoming/outgoing arcs.
Ultimately, the overall procedure corresponds to  extracting what in the literature is referred to as the $(1,1)$-\emph{D-core} ~\cite{Giatsidis2011} of the input \bmultigraph.

Another general preprocessing consists in preventively filtering nodes based on their values of $\bala$ and  $\balr$.
Specifically, given an \bmultigraph $\G = (\V, \E, w)$ and a node $u \in \V$, the difference in $\balr(u)$ induced by any solution to \staticbalance is lower-bounded and upper-bounded by
$LB(u) = \min_{(v,u) \in \E} w(v,u) - \sum_{(u,v) \in \E} w(u,v)$ and 
$UB(u) = \sum_{(v,u) \in \E} w(v,u) - \min_{(u,v) \in \E} w(u,v)$, respectively.
Therefore, a node $u \in \V$ cannot be part of any solution to \staticbalance (and, thus, it can safely be discarded) if $UB(u) \leq \lb(u) -\bala(u)$, or $LB(u) \geq \ub(u) - \balr(u)$ 

As far as the search-space exploration in \bbalg (Section~\ref{sec:bbalg}), we process the arcs in non-increasing amount order.
The reason is that arcs with larger amount are likely to contribute more to the optimal solution.
In terms of visiting strategy, we experimented with both \textsc{bfs} and \textsc{dfs}, observing no substantial difference between the two.

Finally, in  \bsalg (Algorithm~\ref{alg:bsalg}) we compute the set $\cycles'_2$ of (admissible) cycle pairs (Line~5) as follows.
Let $\cycles'$ (Line~4) be partitioned into $\cycles'_{adm}$ (admissible cycles) and $\cycles'_{\neg adm}$ (non-admissible cycles).
For all $C \in \cycles'$, let $\cycles'_{adm}(C) = \{ C' \in \cycles'_+ \setminus \{C\} \mid C \cap C' \neq \emptyset\}$ and $\cycles'_{\neg adm}(C) = \{ C' \in \cycles'_{\neg adm} \setminus \{C\} \mid C \cap C' \neq \emptyset\}$. The set $\cycles'_2$ is computed as (starting with $\cycles'_2 \!=\! \emptyset$):
{
\begin{algorithmic}
\small
\FORALL{$C \in \cycles'_{adm}$}
	\FORALL{$C' \!\in\! \cycles'_{adm}(C) \!\cup\! \cycles'_{\neg adm}(C)$  s.t. $C \!\cup\! C'$ is admissible}
		\STATE $\cycles'_2 \gets \cycles'_2 \cup \{\{C, C'\}\}$
	\ENDFOR
	\FORALL{$C' \in \cycles'_{adm} \setminus \cycles'_{adm}(C)$ s.t. $C \cup C'$ is admissible}
		\STATE $\cycles'_2 \gets \cycles'_2 \cup \{\{C, C'\}\}$
	\ENDFOR
\ENDFOR
\FORALL{$C \in \cycles'_{\neg adm}$}
	\FORALL{$C' \!\in\! \cycles'_{adm}(C) \!\cup\! \cycles'_{\neg adm}(C)$  s.t. $C \!\cup\! C'$ is admissible}
		\STATE $\cycles'_2 \gets \cycles'_2 \cup \{\{C, C'\}\}$
	\ENDFOR
\ENDFOR
\end{algorithmic}
}
\noindent
The rationale is that, for any $C \in \cycles'_{adm}$, there is no need to consider any $C' \in \cycles'_{\neg adm} \setminus \cycles'_{\neg adm}(C)$, as an admissible cycle $C$ cannot be admissible if coupled with a non-admissible cycle that does not overlap with $C$.
Similarly, for any $C \in \cycles'_{\neg adm}$, there is no need to consider any $C' \in \cycles'_{adm} \setminus \cycles'_{adm}(C)$ or any $C' \in \cycles'_{\neg adm} \setminus \cycles'_{\neg adm}(C)$, as a non-admissible cycle $C$ cannot become admissible if coupled with any other cycle that does not overlap with $C$.

\section{Experiments}
\label{sec:experiments}

%

\spara{Datasets.}
We tested the performance of our algorithms on a random sample of two \emph{real datasets} provided by \unicredit, a noteworthy pan-European commercial bank.
The first sample (\firstlog) spans one year in \mbox{2015-16}, while the second, more recent sample (\secondlog) spans one year in \mbox{2017-18}.
Both datasets roughly include 5M receivables and 400K customers.
\revision{Obviously, we worked on an anonymized version of the logs: the datasets contain sensitive information -- such as the identity of the customers, and other personal data -- that cannot be publicly disclosed, and was also made inaccessible to us.}


\spara{Customers' attributes.} 
We set customers' attributes by computing statistics on a training prefix of 3 months, 

The initial actual balance $\bala(u)$ of a customer $u$ was set equal to the average absolute daily difference between the total amount of her passive receivables and the total amount of her active receivables, 
computed over the days when such a difference yielded a negative value. The rationale is that in those days the customer would have needed further liquidity in addition to that provided by incoming receivables, to finalize the payment 
of the receivables where she acted as the debtor. Based on the assumption that real customers would try a new service with a limited initial cash deposit, we also imposed an initial upper bound of $50K$ euros on the actual balance of customers. 

We set the  $\ub$ of a customer to be either  ($i$) \emph{finite}, and, specifically, equal to her average daily incoming amount in the training data (using the average $\ub$ of all customers as the default value for customers with no incoming payments in the training interval), 
or 
($ii$) $\ub = \infty$: here, accounts were allowed to grow arbitrarily.

Finally, we set $\lb(u) = 0$, for each customer $u$.
\revision{The heuristics we use are the result of several discussions with marketing experts of our partner institution. }


\spara{Simulation.}
We defined 6 simulation settings:
\vspace{-0mm}
\begin{enumerate}
\item \emph{Finite CAP.} Let $fcap(u)$ be the finite value of the $\ub$ of a customer $u$ computed as above. We considered 3 scenarios:
\begin{itemize}
\item \emph{\!\!Worst:} $\lifetime(R) \!= \!5, \forall R \!\in\! \mathcal{R};  \ub(u) \!=\! fcap(u), \!\forall u \!\in\! \mathcal{U}$;
\item \mbox{\emph{\!\!Normal:} $\lifetime(R) \!=\! 10, \forall R \!\in\! \mathcal{R};  \ub(u) \!=\! 2fcap(u), \!\forall u \!\in\! \mathcal{U}$;}
\item \emph{\!\!Best:} $\lifetime(R) \!=\! 15, \forall R \!\in\! \mathcal{R};   \ub(u) \!=\! 3fcap(u), \!\forall u \!\in\! \mathcal{U}$;
\end{itemize}
\vspace{1mm}
\item \emph{CAP $= \infty$.} We considered \emph{Worst}, \emph{Normal}, and  \emph{Best} scenarios here too, with \lifetime values equal to  the corresponding 
finite-CAP cases, but we set $\ub(u) = \infty$, $\forall u \in \mathcal{U}$.
\end{enumerate}
\vspace{-0mm}

Such settings identify different sets of valid receivables for a day, and yield different multigraphs.
Table~\ref{tbl:graph-stats} reports on the sizes of the multigraphs extracted from the selected datasets (by using the \hybridalg algorithm to settle receivables at any day of the simulation).
It can be observed that more complex scenarios clearly lead to larger graphs. 
Conversely,  as the $\ub$ goes from $<\!\!\infty$ to  $\infty$ (in the same scenario), graphs get smaller. 
This is motivated as $\ub\!<\!\infty$ represents a tighter constraint for receivable settlement, i.e., more receivables not settled any day which will be included in the input graph of the next day.


\spara{Assessment criteria.}
We measure the performance of the algorithms in terms of the total amount of the receivables selected for settlement.
This metric provides direct evidence of the benefits for both the funder and the customers: the greater the amount, the less liquidity the funder has to anticipate, and the smaller the fees for customers. 

\spara{Parameters.}
Unless otherwise specified, all experiments refer to $\maxcyclelength=15$ (maximum length of a cycle), $L_p=15$ (maximum length of a path between cycles), $H=20$ (size of a connected component to be handled with the exact \bbalg algorithm), and  $K=K_p=1\,000$ (size of the subset of cycles to be used in every iteration of the \bsalg algorithm and the \pathalg algorithm, respectively). 
\revision{
These values were chosen \emph{experimentally}, i.e., by verifying the limits that our implementation could handle, with a good tradeoff between effectiveness and efficiency.
}

\spara{Testing environment.}
All algorithms were implemented in \textsf{Scala} (v. 2.12).
Experiments were run on an i9 Intel 7900x 3.3GHz, 128GB RAM machine.

\subsection{Results on \firstlog}

Table \ref{tbl:experimental-results} shows the results of our experiments on the \firstlog dataset, 
where we compared our ultimate proposal, i.e., \hybridalg (Algorithm~\ref{alg:hybridalg}), against the simple greedy-cycle-selection \johnson algorithm (Algorithm~\ref{alg:johnson}), and the beam-search \bsalg algorithm (Algorithm~\ref{alg:bsalg}).
Clearly, the exact \bbalg (Algorithm~\ref{alg:staticbalanceBB}) could not afford the size of the real graphs involved in our experiments.
However, we recall that it is part of \hybridalg, where it is employed to handle the smaller connected components. To get an idea of its performance, one can thus resort to the comparison  \hybridalg vs. \bsalg.

For each scenario, we split the dataset into 3-month periods, to have a better understanding of the performance on a quarterly basis, 
and speed up the evaluation by parallelizing the computations on different quarters.
For each scenario/quarter pair, we report: total amount (euros) of the settled receivables, 
per-day running time (seconds) averaged across all days in the quarter, total number of settled receivables, and number of distinct customers involved in at least a daily solution. 
For \hybridalg we also report the percentage gain on the total amount with respect to the competitors.

\hybridalg outperforms both the other methods in terms of settled amount.
In the finite-\ub, worst scenario,
the gain of \hybridalg over \johnson is $150\%$ on average, with a maximum of $460\%$.
As for infinite \ub, \hybridalg yields avg gain over \johnson of $70\%$, $28\%$, and $30.6\%$ in the three scenarios.
%
The superiority of \hybridalg is confirmed over the other competing method, \bsalg: the average gain is 
$139\%$, $ 45\%$, and $33\%$  in the finite-\ub scenarios, and $37\%$, $16\%$, and $9\%$ in the infinite-\ub cases, with maximum gain of $392\%$. 
\emph{This attests the relevance of employing the exact algorithm even only on small components.}
In one quarter \bsalg and \hybridalg perform the same: the corresponding graph has no small components where the exact solution improves upon \bsalg.

\begin{table*}[t]
\vspace{-0mm}
\begin{center}
{
\captionsetup{justification=justified,singlelinecheck=off,font={stretch=0.8}}
\caption{{\footnotesize \label{tbl:experimental-results} Experimental results: proposed \hybridalg algorithm  vs. simpler methods.
\textbf{\#R}: total number of
settled receivables. \textbf{\#C}: number of distinct customers involved in at least a daily solution.
\firstlog dataset.}}
\vspace{-0.2cm}
\scriptsize
\begin{tabular}{@{}l@{ ~ }l || S[table-format=9]@{    }S[table-format=3.2]@{    }S[table-format=5]@{  ~  }S[table-format=4]@{    } | S[table-format=10]@{    }S[table-format=4.2]@{  ~ }S[table-format=5]@{ ~   }S[table-format=4]@{    } || S[table-format=10]@{ }S[table-format=3.2]@{ }S[table-format=3.2]@{ }S[table-format=4.2]@{  ~  }S[table-format=5]@{  ~  }S[table-format=4]@{    }} 
 & &   \multicolumn{4}{@{}c@{  }}{\textbf{\johnson}} & \multicolumn{4}{@{}c@{  }}{\textbf{\bsalg}} & \multicolumn{6}{@{}c@{  }}{\textbf{\hybridalg}} \\
\textbf{sce-}  & \multicolumn{1}{@{  }c@{  }||}{\textbf{quar-}}  &   {\multirow{2}{*}{\textbf{amount}}} &  {\multirow{2}{*}{\textbf{time (s)}}} & {\multirow{2}{*}{\textbf{\#R}}} & {\multirow{2}{*}{\textbf{\#C}}} & {\multirow{2}{*}{\textbf{amount}}} &   {\multirow{2}{*}{\textbf{time (s)}}} & {\multirow{2}{*}{\textbf{\#R}}} & {\multirow{2}{*}{\textbf{\#C}}} & {\multirow{2}{*}{\textbf{amount}}} & \multicolumn{1}{@{ }c@{ }}{\textbf{\%gain vs.}}  &  \multicolumn{1}{@{ }c@{ }}{\textbf{\%gain vs.}}  &  {\multirow{2}{*}{\textbf{time (s)}}} & {\multirow{2}{*}{\textbf{\#R}}} & {\multirow{2}{*}{\textbf{\#C}}} \\ 
\textbf{nario} &  \multicolumn{1}{@{  }c@{  }||}{\textbf{ter}} & &  &  & & &  & & & &  \multicolumn{1}{@{ }c@{ }}{\textbf{\textsf{S}-\textsc{bb-lb}}} & \multicolumn{1}{@{ }c@{ }}{\textbf{\textsf{S}-\textsc{beam}}} &  & & \\
\hline
\multirow{2}{*}{\textbf{W}} & 
$Q_1$ &  95397950  & 1.78	 & 1827 & 859 & 79014431   & 3.93 & 2532 & 939 & \bfseries 146703722	&  53.78 & 85.67 & 3.54 &  3330  & 1208 \\
 &  $Q_2$ &   121680562 &  1.80 & 1798 & 863 & 111184064  & 9.22 &  2222 & 891 & \bfseries 157007283	&  29.03 & 41.21 & 9.45 &  3263  & 1248 \\
\textbf{cap:}  &   $Q_3$ &  52232665  & 1.93	 & 2066 & 987 & 59767431  & 14.39 &  2730 & 1034 & \bfseries 82875118	&  58.67 & 38.66 & 14.47 &  3619 & 1318 \\
$< \infty$  &  $Q_4$ &  46457493  &	 1.93	 & 1862 & 950 & 52872191  & 16.49 & 2499 & 987 & \bfseries 260196811 &  460.08  & 392.12 & 13.72 &  3296 & 1254 \\
 \hline
\multirow{2}{*}{\textbf{N}} & 
$Q_1$ &  162215076 &	2.73 &  4535  & 1986  & 146759007 & 99.09  & 6673  & 2269 & \bfseries 221401342 &  36.49 & 50.86 & 75.18  & 7951  &  2685 \\
 &  $Q_2$ & 151707991 & 2.83 &  4305  & 1975 & 133556473 & 47.52  & 5906  & 2141  & \bfseries 168556675 &  11.11 & 26.21 & 48.14  & 7080  &  2548 \\
\textbf{cap:} &  $Q_3$ &  143547779 &	3.01 &  5093  & 2200 & 158408830 & 76.15  & 7319  & 2464  & \bfseries 208011456 &  44.91 & 31.31 & 113.72 & 8172  &  2730 \\
$< \infty$ &  $Q_4$  &  149813738 &	 3.01 &  5137  & 2296 & 162254519 & 104.73  & 7459  & 2603 & \bfseries 277568569 &  85.28 & 71.07 & 117.45  & 8551  &  2941 \\
 \hline
\multirow{2}{*}{\textbf{B}} & 
$Q_1$  &   229604568	 & 7.63  &  7443  & 3115 & 263379260	& 172.09	& 11045	& 3566 & \bfseries 270221655 &  17.69	& 2.60 & 158.36 & 12004 & 3957	 \\
 &  $Q_2$ &   195171304	& 4.00  & 7183	 & 3108 & 208162233	& 133.60	& 10364	& 3506 & \bfseries 251353161 &  28.79	& 20.75 & 143.71 & 11669 & 3917	\\
\textbf{cap:} &  $Q_3$  &   236195806	 & 9.73  & 8397 & 3516 & 264654596	& 193.91	& 12564	& 4000 & \bfseries 314861548 &  33.31	& 18.97 & 205.36 & 13486 & 4333	\\
$< \infty$ &  $Q_4$ &   210056244	 & 8.92  & 9094 & 3801 & 234928905	& 212.45	& 13184	& 4259 & \bfseries 449862686 &  114.16	 & 91.49 & 239.25	 & 14360 & 4620 \\
 \hline
 \hline
\multirow{2}{*}{\textbf{W}} & 
$Q_1$ &  194399317	 & 2.56 & 2863	 & 1275 &  215249395	 	& 110.43	 & 5418	 & 1535	&  \bfseries 295595794	&  52.06 & 37.33 & 	143.39 & 6528	 & 1872 \\	
&  $Q_2$  &   215635790	 & 2.30 & 2939	 & 1283 &  263672222		& 124.54	 & 5285	 & 1490	&  \bfseries 319804764	&  48.31 & 21.29 & 	125.16 & 6724	 & 1918 \\
\textbf{cap:} &  $Q_3$  &   206616203	 & 12.06 & 3329 & 1371 & 262778883		& 343.81	 & 6876	 & 1718	&  \bfseries 302740733	&  46.52 & 15.21 & 	 354.13 & 8290 & 2116 \\
$\infty$ &  $Q_4$ &  216264862	 & 2.56 & 3224	 & 1383 &  286859858	 	& 287.28	 & 6034	 & 1685	&  \bfseries 504402450	&  133.23 & 75.84 &  304.92 & 7241 & 1992	\\
 \hline
\multirow{2}{*}{\textbf{N}} & 
$Q_1$  &  553364544	 & 64.58	& 7131 & 2836	&  660907304	 	& 1168.95	 & 15323	& 3761	&  \bfseries 779733449	&   40.91	& 17.98 &  1006.17 &  17082	 & 4268	\\
&  $Q_2$ &  643722123	 & 6.67 & 6742	 & 2736	&  663873349		& 618.98	& 14570	& 3507	&  \bfseries 784314578	&   21.84	& 18.14 &  690.14	 &  16761	& 4133	\\
\textbf{cap:} &  $Q_3$ &  693852990	& 29.03 & 7999 & 3034	&  743945529		& 1159.27	 & 17390	& 4143	&  \bfseries 827346450	&   19.24	& 11.21 &  1329.80 &  19544	 & 4701	\\
$\infty$ &  $Q_4$ &  751368135	& 30.81 & 8289 & 3189	&  855932063	 	 & 757.85	 & 17666	&  4155	&  \bfseries 987866224	&   31.48	& 15.41 &  865.92	 &  19576	 & 4718	\\ \hline
\multirow{2}{*}{\textbf{B}} & 
$Q_1$ &    916036152	& 8.35	& 11246	& 4231 &  1110498564		& 866.00	& 23055	& 5353 & \bfseries 1172926462	&  28.04 & 5.62 &	 988.04	& 24811	& 5917	\\
&  $Q_2$ &    1028777612	 & 26.28	& 10937	& 4159 &  1275000083		& 842.56	& 23235	& 5333 & \bfseries 1565296054	&  52.15 & 22.77 & 1182.68	 & 25469	& 5901	\\
\textbf{cap:} &  $Q_3$ &   1329747599	 & 54.87	 & 13271	& 4830 &  \bfseries 1489713871		& 993.90	& 27404	& 6130  & \bfseries 1489713871	&  12.03 & 0 & 1101.04	& 27404	& 6130	\\
$\infty$ &  $Q_4$ &  1270225872	 & 22.05	& 13337	& 4835 &  1524865674		& 904.29	& 26177	& 5781 & \bfseries 1657784888 &  30.51 & 8.72 &  886.05	& 27746	& 6239	\\
\end{tabular}}
\end{center}
\vspace{-2mm}
\end{table*}

Concerning running times, the fastest method is the simplest one, i.e., \johnson: in the finite-\ub cases it takes on average $10s$. 
The proposed \hybridalg takes from $3.5s$ to $22mins$, remaining perfectly compliant with the settings 
of the service being built, which computes solutions offline, at the end of each day. 
\hybridalg is comparable to \bsalg, due to the fact that the computation
on large components dominates. 

\spara{Scalability.}
To test the scalability of \hybridalg, we considered the finite-cap, normal scenario and randomly sampled segments of data corresponding 
to $5$, $10$, $15$, $30$, $60$, and $90$ days. We collapsed the set of receivables of each segment into one single \bmultigraph, and ran \hybridalg on it, setting $\maxcyclelength = 10$ and $K = 100$. 
The goal of this experiment was to assess the running time of \hybridalg on larger graphs. 
Table \ref{table:scalability} reports the outcome of this experiment, showing, for each segment, size of the \bmultigraph, settled amount, and running time (in seconds).
Note that amounts in Table~\ref{tbl:experimental-results} 
are generally larger than those reported here. 
The two experiments are not comparable in terms of amount, as Table~\ref{tbl:experimental-results} 
reports amounts summed over all time instants of a quarter, each expanded to a window of 5, 10 or 15 days depending on the \lifetime parameter. 
Running time is instead comparable up to the segments of 15 days, because the first table reports average per-day running time.
Indeed, when up to $15$ days are considered, the achieved times are consistent with those reported before. 
On a period of $30$ days, the algorithm is still very efficient, taking slightly more than a minute. 
The running time increases on the two larger segments: the algorithm employs less than one hour on the two-month segment, and around $7.5$ hours on the three-month one. 
Albeit the cost increase is remarkable, note that the network-based RF service is designed to function offline, at the end of each working day.
In this setting even the larger times reported would be acceptable.
Moreover, given that the service works on a daily basis, the realistic data sizes that \hybridalg is required to handle, are those of the previous experiment.
Finally, the tested implementation is sequential. It can be improved by parallelizing the computation on the connected components of the input multigraph.

\begin{table}[t]
\vspace{-0mm}
\begin{center}
\scriptsize
\caption{\label{table:scalability}{\footnotesize Scalability of the proposed \hybridalg algorithm \\ (\firstlog dataset).}}
\vspace{-1mm}
\begin{tabular}{S[table-format=2]@{ }S[table-format=6]@{      }S[table-format=7]@{      }S[table-format=9]@{ }S[table-format=5]@{}}
{\bf days} & {\bf nodes} & {\bf arcs} & {\bf amount} & {\bf time (s)} \\
\hline
5 & 15983 & 14466 & 185959 & 1 \\ 
10 & 41088 & 43244 & 873317 & 4\\
15 & 68183 & 85454 & 3471960 & 17 \\
30 & 106167 & 183570 & 16151068 & 65\\
60 & 143989 & 377635 & 38063145 & 3291 \\
90 & 168861 & 600172 &  73101255 & 27504\\
\hline
\vspace{-4mm}
\end{tabular}
\end{center}
\end{table}

\subsection{Results on \secondlog}

In the following we discuss the results obtained on the \secondlog dataset, which we used to investigate aspects that were not part of the previous evaluation, namely comparison against a baseline, and assessment of the path-selection methodology and the approaches for avoiding temporary constraint violation. 
In the following sets of experiments we picked \hybridalg as our reference algorithm, as the previous assessment (on \firstlog) showed that it corresponds to our best cycle-selection-based method.

\vspace{0mm}
\spara{Comparison against a baseline.}
Whilst we are not aware of any external method, as this is (to the best of our knowledge) the first attempt to exploit the receivable network to optimize RF, we define a possible baseline for the proposed algorithm(s) as follows.
We start from a solution $\E^* = \E$ composed of all the arcs of the input multigraph, and, as long as $\E^*$ violates some Problem~\ref{problem:balancestatic}'s constraints, we iteratively:
($i$) remove arcs from $\E^*$ in non-decreasing amount order, until Constraint~(\ref{eq:prob1constraint1}) in Problem~\ref{problem:balancestatic} is satisfied; 
($ii$) extract the $(1,1)$-D-core from the subgraph induced by $\E^*$ (to satisfy Constraint~(\ref{eq:prob1constraint2}) in Problem~\ref{problem:balancestatic}). 
In Table~\ref{tbl:baseline} we compare this baseline, dubbed \rfbaseline, and \hybridalg. 
We split the \secondlog dataset into 3-month periods, to have an understanding of the performance on a quarterly basis, 
and speed up the evaluation by parallelizing on different quarters.
\revision{For each quarter, parameter-configuration, and algorithm, we report: total
amount (euros) of the settled receivables, number of settled receivables (\#R), and number of distinct customers involved in a daily solution (\#C).}
The main observation here is that the baseline achieves a consistent loss ($70\%$--$100\%$) against \hybridalg in all quarters and scenarios: this confirms the strength of the proposed algorithm.

\begin{table}[t]
\vspace{-0mm}
\captionsetup{justification=justified,singlelinecheck=off,font={stretch=0.8}}
\begin{center}{
\revision{
\caption{\revision{\footnotesize \label{tbl:baseline} Comparing best cycle-based algorithm (\hybridalg) vs. \rfbaseline baseline in terms of (1)  total amount (euros) of settled receivables, (2) total number of
settled receivables (\textbf{\#R}), and (3) number of distinct customers involved in at least a daily solution (\textbf{\#C}). \secondlog dataset.}}
\vspace{-0.1cm}
\scriptsize
\begin{tabular}{l  |  l  || S[table-format=8] S[table-format=4] S[table-format=3] ||  S[table-format=8] S[table-format=1] S[table-format=1]  }  
\multicolumn{1}{l|}{\multirow{2}{*}{\textbf{quar-}}} & \multicolumn{1}{l||}{\multirow{2}{*}{\textbf{sce-}}} & \multicolumn{3}{c||}{\textbf{\hybridalg}} & \multicolumn{3}{c}{\textbf{\rfbaseline}}  \\
\multicolumn{1}{l|}{\textbf{ ter}} & \multicolumn{1}{l||}{\textbf{nario}}  & \multicolumn{1}{c|}{\textbf{amount}} & \multicolumn{1}{c|}{\textbf{\#R}} & \multicolumn{1}{c||}{\textbf{\#C}}  & \multicolumn{1}{c|}{\textbf{amount}} & \multicolumn{1}{c|}{\textbf{\#R}} & \multicolumn{1}{c}{\textbf{\#C}} \\  
\hline
 $Q_1$ & {\multirow{2}{*}{\textbf{W}}} & 208452169  & 2045 & 853 &  75871408 & 8 & 6\\ 
 $Q_2$ & & 133998727 & 2482 & 933 &  28759273 & 4 & 3 \\ 
 $Q_3$ & \textbf{cap:}  & 250919555 & 2445 & 1007&  57472835 & 4 & 3 \\ 
 $Q_4$ &  {\multirow{1}{*}{\textbf{$<\!\!\infty$}}}  & 435864820 & 2678 & 1019 & 128409243 & 8 & 3 \\ 
\cline{1-2}\cline{3-5}
$Q_1$ & {\multirow{2}{*}{\textbf{N}}} &  100052350 & 3582 & 1578 & 0  & 0 & 0 \\ 
$Q_2$ & &  210803336 & 5054 & 1832 & 0 & 0 & 0 \\ 
$Q_3$ &  \textbf{cap:} &  181024199 & 5069 & 2083& 0 & 0 & 0 \\ 
$Q_4$ & {\multirow{1}{*}{\textbf{$<\!\!\infty$}}} & 405407829 & 5171 & 1998 & 68940677 & 4 & 3\\ 
\cline{1-2}\cline{3-5}
 $Q_1$  & {\multirow{2}{*}{\textbf{B}}} &  128172168  & 5369 & 2347 & 0 & 0 & 0 \\ 
 $Q_2$  &   & 223972720 & 7429 & 2740 & 0 & 0 & 0 \\ 
 $Q_3$  & \textbf{cap:} &  138227899 & 7934 & 3262& 0 & 0 & 0 \\ 
 $Q_4$  &  {\multirow{1}{*}{\textbf{$<\!\!\infty$}}}  & 380133762 & 7786 & 2968 & 0 & 0 & 0 \\ 
 \hline
 \hline
 $Q_1$ & {\multirow{2}{*}{\textbf{W}}} & 250919555 & 4474 & 1504 & 75871408  & 8 & 6 \\ 
 $Q_2$ &  &  374766629 & 5194 & 1525 &  28759273  & 4 & 3 \\   
 $Q_3$ & \textbf{cap:} &  300021017 & 5321 & 1603 & 57472835 & 4 & 3 \\  
 $Q_4$ &  {\multirow{1}{*}{\textbf{$\infty$}}}  &  516434249 & 6896 & 1645 & 128409243 & 8 & 3 \\  
\cline{1-2}\cline{3-5}
$Q_1$ & {\multirow{2}{*}{\textbf{N}}}   & 395904601 & 9373 & 3012 & 0 & 0 & 0 \\ 
$Q_2$ & &   555117105 & 11712 & 3270 &  0 & 0 & 0 \\ 
$Q_3$ & \textbf{cap:} &  471313548 & 14342 & 3888 & 0 &  0 & 0\\ 
$Q_4$ & {\multirow{1}{*}{\textbf{$\infty$}}}  &  679395059 &15319 & 3581 & 189533414 & 26 & 15 \\ 
\cline{1-2}\cline{3-5}
 $Q_1$  & {\multirow{2}{*}{\textbf{B}}}   & 552458593 & 14314 & 4566 & 0  & 0 & 0 \\ 
 $Q_2$  &   & 699245106 & 19557 & 5519 & 0 & 0 & 0 \\ 
 $Q_3$  & \textbf{cap:} & 544101823 & 22306 & 6078 & 0 & 0 & 0  \\ 
 $Q_4$  &   {\multirow{1}{*}{\textbf{$\infty$}}} & 907115932 & 23606 & 5881 & 120592737 & 22 & 15 \\ 
\end{tabular}}
}
\end{center}
\vspace{-2mm}
\end{table}

\begin{table*}[t]
\vspace{-0mm}
\begin{center}{
\caption{\footnotesize \label{tbl:path-results} Evaluating path selection: comparing the best cycle-based algorithm, \hybridalg, to the two path-based algorithms (\secondlog dataset). 
}
\vspace{-0.2cm}
\scriptsize
\begin{tabular}{ c c c  | S[table-format=4]   S[table-format=9]  | S[table-format=4]     S[table-format=9]   S[table-format=4]     | S[table-format=4]   S[table-format=9]     S[table-format=4]     S[table-format=2.1]     } 
\multicolumn{1}{c}{}  & \multicolumn{1}{c}{}  &  \multicolumn{1}{c}{} & \multicolumn{2}{|c}{\textbf{\hybridalg}} & \multicolumn{3}{|c}{\textbf{\pathalggreedy}} & \multicolumn{4}{|c}{\textbf{\pathalgsuk}} \\
\multicolumn{1}{c}{\multirow{2}{*}{\textbf{cap}}} & \multicolumn{1}{c}{\textbf{sce-}} & \multicolumn{1}{c}{\textbf{quar-}}  & \multicolumn{1}{|c}{\textbf{time}}  & \multicolumn{1}{c} {\multirow{2}{*}{\textbf{amount}}} &  \multicolumn{1}{|c}{\textbf{time}}  &  {\multirow{2}{*}{\textbf{amount}}} & \multicolumn{1}{c}{\textbf{\%gain vs}}  &  \multicolumn{1}{|c}{\textbf{time}}  &  {\multirow{2}{*}{\textbf{amount}}} & \multicolumn{1}{c}{\textbf{\%gain vs}}  &  \multicolumn{1}{c}{\textbf{\%gain vs}}  \\
\multicolumn{1}{c}{} & \multicolumn{1}{c}{\textbf{nario}}  &  \multicolumn{1}{c}{\textbf{ter}} &   \multicolumn{1}{|c}{\textbf{(s)}} & &  \multicolumn{1}{|c}{\textbf{(s)}} & &  \multicolumn{1}{c}{\textbf{\hybridalg}}  &  \multicolumn{1}{|c}{\textbf{(s)}} & &  \multicolumn{1}{c}{ \textbf{\pathalggreedy}} &   \multicolumn{1}{c}{\textbf{\hybridalg}}  \\ 
\hline
{\multirow{12}{*}{\textbf{$<\!\!\infty$}}} & \multirow{4}{*}{\textbf{W}} &  $Q_1$  & 4 & 208452169 & 4 & 206122369 & -1 &  54 & 208658217	& 1 & 0.1 \\
& &    $Q_2$  & 6 & 133998727 &  6 & 142880339 & 7 & 38 & 139157842	& -3 & 3.9   \\
& &  $Q_3$  & 5 & 250919555 & 4 & 251236654 & 0 & 18	& 251689885	& 0 & 0.3  \\
& &   $Q_4$  & 6 & 435864820 & 6 & 436866648 & 0 & 54	& 437305782	& 0 & 0.3  \\
\cline{2-12}
& \multirow{4}{*}{\textbf{N}} &   $Q_1$  & 8 & 100052350 & 8 & 171073852 & 71 & 375 & 130166286 & -24 & 30.1  \\
& &  $Q_2$  & 95 & 210803336 & 47 & 224572822 & 7 & 525	& 222848364	& -1 & 5.7  \\
&  &  $Q_3$  & 13 &  181024199 & 14 & 187757266 & 4 & 429	& 215370572	&15 & 19.0  \\
& &  $Q_4$  & 23 & 405407829 & 38 & 425474537 & 5 & 448 & 413760885	& -3 & 2.1  \\
 \cline{2-12}
& \multirow{4}{*}{\textbf{B}} &  $Q_1$  & 50 & 128172168 &  47 & 139304792 & 9 & 576	& 140389844	& 1 & 9.5  \\
&  &   $Q_2$  & 215 & 223972720 & 185 & 238760149 & 7 & 721 & 245781771	& 3 & 9.7  \\
&  & $Q_3$  & 66 & 138227899 & 34 & 154369240 & 12 & 677	& 158211843 & 2 & 14.5 \\
&  & $Q_4$  & 192 & 380133762 & 109 & 395718595 & 4 & 1150 & 392670886 & -1 & 3.3  \\
\hline
\hline
{\multirow{12}{*}{\textbf{$\infty$}}} & \multirow{4}{*}{\textbf{W}} &  $Q_1$   & 20 & 374766629 &  7 & 401557792 & 7 &  256	& 407383061 & 1 & 8.7  \\
 & &  $Q_2$  & 169 & 226076200 & 122 & 250268430 & 11 & 330	& 264854214 & 6 & 17.2  \\
& & $Q_3$  & 17 & 300021017 & 17 & 317522710 & 6 & 300 & 316476954	& 0 & 5.5  \\
& &  $Q_4$   & 98 & 516434249 & 101 & 540613736 & 5 & 214 & 543879376	& 1 & 5.3  \\
\cline{2-12}
& \multirow{4}{*}{\textbf{N}}  &  $Q_1$  & 246 & 395904601 & 162 & 434561172 & 10 &  1761 & 458001083 & 5 & 15.7  \\
&  & $Q_2$  & 968 & 555117105 & 739 & 622621171 & 12 &   2023 & 624403545 & 0 & 12.5 \\
&  &  $Q_3$  & 573 & 471313548 & 505 & 516884575 & 10 &  1766 & 604594618 & 17 & 28.3  \\
&  & $Q_4$  & 884 & 679395059 & 694 & 713902252 & 5  & 2279 & 727907790 & 2 & 7.1  \\
 \cline{2-12}
& \multirow{4}{*}{\textbf{B}} &  $Q_1$  & 804 & 552458593 & 634 & 602544685 & 9 & 3004 & 538053935  & -11 & -2.6  \\
&  & $Q_2$  & 1604 & 699245106 & 1170 & 819639994 & 17 &  4822 & 727206932 & -11 & 4  \\
& &  $Q_3$  & 971 & 544101823 & 814 & 647240125 & 19 &  3430 & 685719252 & 6 & 26  \\
&  & $Q_4$  & 1609 & 907115932 & 943	& 970525923 & 7 &  3360 & 963608594	& -1 & 6.2  \\
\end{tabular}}
\end{center}
\vspace{-4mm}
\end{table*}

\begin{table}[t]
\vspace{-1mm}
\begin{center}{
\caption{\footnotesize \label{tbl:tmp_violations} Evaluating method for avoiding temporary constraint violation (Algorithm: \hybridalg; $\ub <\infty$; \secondlog dataset)}
\vspace{-0.1cm}
\scriptsize
\begin{tabular}{ c  |  cS[table-format=8] } 
\multicolumn{1}{c|}{\textbf{quarter}} & \multicolumn{1}{c}{\textbf{scenario}} & \multicolumn{1}{c}{\textbf{amount}}  \\
\hline
 $Q_1$  & \multirow{4}{*}{\textbf{W}} &  7836382 \\
$Q_2$ &   &  10821713    \\
$Q_3$  &  &  11026866  \\
 $Q_4$  &    & 18073953  \\
 \hline
 $Q_1$  & \multirow{4}{*}{\textbf{N}} &  7836382  \\
$Q_2$. &   &  10821713    \\
$Q_3$  &  &  11026866  \\
 $Q_4$  &    & 18073953   \\
 \hline
 $Q_1$  & \multirow{4}{*}{\textbf{B}} &  7836382  \\
$Q_2$. &   &  10821713   \\
$Q_3$  &  &  11026866  \\
 $Q_4$  &    & 18073953  
\end{tabular}}
\end{center}
\vspace{-7mm}
\end{table}

\spara{Evaluating path selection.}
%
A further experiment we carried out was on the impact of enriching a cycle-based solution by adding paths between cycles. 
To this end, we compared \hybridalg to its path-selection version \pathalg.
Specifically, we consider the two versions of \pathalg mentioned in Section~\ref{algo:paths}: 
\pathalggreedy, which performs a simple greedy path selection, and, \pathalgsuk, which employs the more refined selection method in Algorithm~\ref{alg:bsalg}.
Both variants are based on the outline in Algorithm~\ref{alg:pathalg}.
The results of this experiment are  in Table~\ref{tbl:path-results}.

The main finding here is that our path-based algorithms outperform the best no-path method in all configurations (but one, where they however exhibit small losses, i.e., 1\% and 2.6\%).
\pathalggreedy achieves an average gain over \hybridalg of, respectively, $1.5\%$, $21.8\%$ and $8\%$ in the finite-cap worst, normal, and best scenarios, and $7.3\%$, $9.3\%$, and $13\%$ in the infinite-cap worst, normal, and best cases. 
The average gain of \pathalgsuk over \hybridalg is, respectively, $1.2\%$, $14.2\%$, and $9.2\%$ in the finite-cap worst, normal, and best scenarios, and $9.2\%$, $15.9\%$, and $8.4\%$ in the infinite-cap worst, normal, and best cases.
\emph{This attests the relevance of the idea of adding paths to the cycle-based solutions}. 

The two path-selection strategies are comparable: 
\pathalgsuk wins in 12 configurations, with 5\% avg gain, while \pathalggreedy wins in 8 configurations, with 7\%~avg~gain.

As for running time, \pathalggreedy is  comparable to \hybridalg, and actually faster in $75\%$ of cases.
The motivation may be that adding paths to a daily solution causes the algorithm to work on smaller graphs in the next days.
As expected, \pathalgsuk is instead consistently (about one order of magnitude) slower than \pathalggreedy and \hybridalg, due to its more sophisticated path-selection strategy.

\spara{Avoiding temporary constraint violation.}
We also tested the performance of Algorithm~\ref{alg:negbal} for avoiding temporary constraint violation, 
employing the \hybridalg algorithm and considering the finite-cap scenario.
The results are reported in Table~\ref{tbl:tmp_violations}.
We observe that, although the settled amount is expectedly less than its counterpart where temporary constraint violation is not addressed, this amount remains reasonably large, i.e., in the order of 1M/10M euros.


\section{Related Work}
\label{sec:related}

The \staticbalance problem that we study in this work is a novel contribution of ours~\cite{bordino18network,bordino20advancing}.
\revision{
To the best of our knowledge, \emph{no previous work has ever adopted a similar, network-based formulation, neither for receivable financing, nor for other applications}.

There are some problems falling into the same broad application domain, while still being clearly different from ours.
The main goal of these problems is mainly to predict -- based on historical data -- whether a receivable will be paid, the date of the payment, and late-payment amounts.
}
In this regard,  Zeng\etal~\cite{zeng2008predictive} devise (supervised) machine-learning models for invoice-payment outcomes, enabling customized actions tailored for invoices or customers. 
Kim\etal~ \cite{kim2016payment} focus on debt collection via call centers, proposing machine-learning models for late-payment prediction and customer-scoring rules 
to assess the payment likelihood and the amount of late payments.
Nanda~\cite{nanda2018proactive} shows that historical data from account receivables helps mitigate the problem of outstanding receivables.
Tater\etal~\cite{tater2018prediction} propose ensemble methods to predict the status of an invoice being affected by other invoices that are simultaneously being processed. 
Cheon~and~Shi~\cite{cheong2018customer} devise a customer-attribute-based neural-network architecture for predicting the customers who will pay their (outstanding) invoices with high probability. 
Appel\etal~\cite{appel2019optimize} present a prototype developed for a multinational bank, aimed to support invoice-payment prediction.

\revision{
It is apparent that all those works do not share any similarity with our network-based formulation of receivable financing: 
our goal is \emph{to select a receivables according to a (novel) combinatorial-optimization problem} that enables money circulation among customers, while the above works \emph{predict future outcomes on the payment} of receivables.
Such prediction problems cannot even be somehow auxiliary for our setting.
In fact, as discussed in Section~\ref{sec:background}, our network-based receivable-financing service \emph{does not allow non-payments by design}.
Thus, in our context, asking whether payments will be accomplished or not is a meaningless question.
}

As far as our algorithmic solutions,  all the references that inspired us are already reported in Section \ref{sec:algo}, where appropriate.
Here, let us just discuss, for the interested reader, a couple of problems that share some marginal similarity with the proposed \staticbalance problem. 
As testified by upper-bound derivation in the design of the exact algorithm, \staticbalance resembles a \emph{network flow} problem~\cite{Ahuja93}.
Among the numerous variants of this problem, the one perhaps closer to ours is 
\textsc{Min-cost-flow-with-minimum-quantities}~\cite{seedig2011flowminq}, which introduces a constraint for having minimum-flow quantities on the arcs of the network. 
This constraint models the fact that in logistics networks one may require that at least a minimum quantity is produced at a site, or nothing at all. 
In our context minimum quantities are required because a receivable can only be paid entirely: fractional payment is not allowed.
Nevertheless, \staticbalance still differs from \textsc{Min-cost-flow-with-minimum-quantities} as it requires a different \emph{conservation-flow} constraint (to properly handle cap and floor), and an additional constraint that each node in the solution has at least one incoming arc and one outgoing arc. 

Another recent network-flow variant is \textsc{Max-flow-problem-with-disjunctive-constraints} \cite{pferschy2013maxflow-conflict}, which introduces binary constraints
on using certain arc pairs in a solution.
The problem is applied to event-participant arrangement optimization~\cite{she2016conflict-aware}  
in event-based social networks, such as \emph{Meetup}. 
Compared to that problem, our \staticbalance requires different constraints on the arcs included in any feasible solution.

\vspace{-0mm}
\section{Conclusion}
\label{sec:conclusion}

We have presented a novel, network-based approach to receivable financing.
Our main contributions consist in a principled formulation and solution of such a novel service. 
We define and characterize  a novel optimization problem on a network of receivables, and design both an exact algorithm, and more efficient, cycle-selection-based algorithms to solve the problem. 
We also improve basic cycle-based methods via proper selection of paths among identified cycles, 
and devise adaptations of our methods to real-world scenarios where temporary constraint violations are not allowed at any time instant.
Experiments on real receivable data attest the good performance of our algorithms.

In the future we plan to incorporate \emph{predictive} aspects in our methods, where recent history is considered instead of taking static decisions every day.
We will also attempt to apply the lessons learned here to advance other financial services, e.g., receivable trading, or dynamic discounting.



\end{document}